\documentclass[preprint,12pt]{elsarticle}
\usepackage[margin=1in]{geometry}
\usepackage{amssymb}
\usepackage{amsmath}
\usepackage{hyperref}
\usepackage{amsthm}
\usepackage[dvipsnames]{xcolor}
\usepackage{subfig}
\usepackage{graphicx}

\journal{}

\begin{document}

\begin{frontmatter}

\title{A Fourier/Modal-Spectral-Element Method for the Simulation of High-Reynolds Number Incompressible Stratified Flows in Domains with a Single Non-Periodic Direction}

\author[1]{Nidia Reyes-Gil}
\author[2]{Greg Thomsen}
\author[3]{Kristopher Rowe} 
\author[1]{Peter Diamessis}

\affiliation[1]{organization={School of Civil and Environmental Engineering},
            addressline={Cornell University}, 
            city={Ithaca},
            state={NY},
            country={USA}}

\affiliation[2]{organization={Wandering Wakhs Research},
            city={Austin},
            state={TX},
            country={USA}}

\affiliation[3]{organization={Leadership Computing Facility, Argonne National Laboratory},
            city={Lemont},
            state={IL},
            country={USA}}

\begin{abstract}
We present the components of a high-order accurate Navier-Stokes solver designed to simulate high-Reynolds-number stratified flows. The proposed numerical model addresses some of the numerical and computational challenges that high-Reynolds-number simulations pose, facilitating the reproduction of stratified turbulent fluid dynamics typically observed in oceanic and atmospheric flows, namely the development of thin regions of high vertical shear, strongly layered turbulence at high Reynolds numbers and internal wave radiation. This Navier-Stokes solver utilizes a Fourier pseudo-spectral method in the horizontal direction and a modal spectral element discretization in the vertical. We adopt an implicit-explicit time discretization scheme that involves solving several one-dimensional Helmholtz problems at each time step. Static condensation and modal boundary-adapted basis functions result in an inexpensive algorithm based on solving many small tridiagonal systems. A series of benchmark studies is presented to demonstrate the robustness of the flow solver. These include two-dimensional and three-dimensional problems, concluding with a turbulent stratified wake generated by a sphere in linear stratification.\\
\end{abstract}


\end{frontmatter}

\section{Introduction}\label{chap:info}
In oceanic and atmospheric flows at environmental scales, the combination of turbulence and density variations introduces additional layers of physical complexity to turbulent dynamics that are not yet fully understood. Turbulence reorganizes under the influence of stratification into coherent patches of vertical vorticity, that interact through focused shearing motions in the vertical and may trigger a full range of potential instabilities and complicated nonlinear dynamics. Furthermore, at high Reynolds numbers,  turbulent fluid motions can persist even longer than in unstratified environments due to thin regions of high shear that can sustain turbulence far after the original turbulent event. In view of the complexity of the phenomenon, this intricate fluid dynamics are increasingly studied through numerical models, given the numerous practical challenges involved in in-situ data acquisition, the wide range of scales that need to be measured and the inherent difficulties to reproduce high-Reynolds-number flows in a stratified environment in the laboratory. 

Numerical models focused on reproducing unsteady turbulent flow through numerical computation, specifically by computing truncated or filtered solutions to the full incompressible Navier-Stokes equations (\cite{geurts2013direct, Domaradzki2002}), can involve significant computational challenges with respect to resource utilization. Their computational limitations, in terms of execution time and memory footprint, depend on the range of leading scales of motion that need to be simulated: computing time is determined by the ratio of the total simulation time to the smallest resolvable time scale (or time step), and memory requirements by the ratio of the domain size to the smallest resolvable length scale \cite{softwareDesign}. In particular, turbulent flows in stratified environments exhibit a broad range of scales in space and time, demanding immense computational resources that make their numerical simulation---accounting for all the length and time scales involved---practically impossible. Simulation strategies such as direct numerical simulations (DNS), large-eddy simulations (LES) \cite{LES_org, Deardorff_1970} and implicit large-eddy simulations (ILES)\cite{Germano86_differentialFilters, BORIS1992199} (see Section \ref{Num_sim_approaches}), have been essential in studying turbulence in stratified environments. However, their main limitation---with respect to resource utilization---has been the range of temporal and spatial scales that can be simulated. 

An example of the limitation above is the numerical simulation of stratified turbulent wakes. Stratified wakes produced by islands, mountains, or seamounts, are a canonical example of stratified turbulent flow which can involve length scales of millimeters to kilometers and time scales from milliseconds to hours, and clearly pose important resource-utilization challenges to be simulated numerically. If we define a geometry based Reynolds number ($Re$), based on the characteristic velocity and length scales of the generating body, i.e. $Re=UD/\nu$, the largest geometry-based $Re$ simulated numerically ($Re \sim O(10^4$--$10^5)$ using ILES \cite{ZhouDiamessis2019}) are still orders of magnitude below values typically observed in geophysical flows in nature ($Re \sim O(10^8$--$10^{10})$) \cite{Brethouwer07}. Efficient computational implementations, combined with high-order numerical models, offer a powerful tool to explore a broader parameter space and gain deeper insights into the complex non-linear dynamics of high $Re$ stratified flows, bridging at least partially the above gap in reference.

As a continuation of \cite{DIAMESSIS2005}, \cite{ZhouDiamessis2019} and \cite{KrisPaper}, this work describes a pseudo-spectral method for simulating the three-dimensional Navier-Stokes equations under the Boussinesq approximation. Particular care is taken to ensure that the model is suitable to simulate stratified turbulent wakes at a geometry-based $Re \sim O(10^5-10^6)$. A defining feature of this work is the application of boundary-adapted polynomial basis functions combined with a domain decomposition approach in the vertical to increase efficiency. The model is well-suited for other stratified, highly turbulent flows developed in long, high aspect-ratio domains, such as shear and boundary layers and jets. To contextualize and delineate the document's objective and motivation, an overview of the key dynamics of stratified turbulent wakes is presented below, followed by a description of the proposed numerical model.

\subsection{Simulation of stratified turbulent flows: studying wakes as test case}
In stratified turbulence, the importance of stratification in the flow dynamics is estimated by comparing kinetic and potential energies. A dimensionless parameter used for such purpose is the Froude number, $Fr = \frac{U}{N l}$, where $l$ and $U$ correspond to appropriate length and velocity scales and $N$ to the stratification frequency or Brunt--V\"{a}is\"{a}l\"{a} frequency. In general, when $Fr\lesssim 1$ stratification effects are considered to be important, and the smaller $Fr$, the larger these effects are \cite{introGFD}. 

Particularly in stratified flows at high $Re$, potential secondary instabilities due to buoyancy-driven shear can appear when the local (turbulent) $Fr$ is less than unity, which takes place at some point during the evolution of a decaying turbulent flow in stable stratification. This occurs during the so-called strongly stratified regime (SSR), where there is still a lack of understanding of turbulent dynamics (see \cite{deBruynKops_Riley_2019, Brethouwer07, Riley2000}). Complex nonlinear motions in the SSR---including internal wave radiation \cite{KrisPaper}, buoyancy-driven shear \cite{dbKAndRiley2003}, and both space-filling and layered turbulence \cite{Portwood_deBKops2016}---significantly influence vertical transport and diapycnal mixing in atmospheric and oceanic flows. 

The wake generated by a sphere towed in a linear stratification profile is a useful canonical model to study complex wake flows and turbulent dynamics in the SSR. A number of the laboratory studies that have addressed the decay of turbulence in stratified flows, particularly in the low $Fr$ number regime, have dealt with towed sphere wakes (see \cite{Lin_Lindberg_Boyer_Fernando_1992, Chomaz93a, bonneton_chomaz_hopfinger_1993,  Riley2000, RILEY2021483}). Results from those experiments can be considered representative of turbulence developed in a stably-stratified fluid, particularly those not sustained by ambient shear flow \cite{RILEY2021483}. In sphere wakes, the means of generating the turbulence is controllable, and the body-based $Re$ and $Fr$ can be easily varied. Moreover, its study is relevant to several geophysical and engineering applications. 

\subsubsection{Simulation strategies} \label{Num_sim_approaches}
Numerical simulation of stratified turbulent flows is achieved by discretizing the incompressible Navier-Stokes equations under the Boussinesq approximation. In principle, it is possible to statistically model fluctuation-correlations using a Reynolds average Navier-Stokes approach. However, far more accurate simulation strategies are needed. Such strategies are increasingly interested in reproducing the main features of the unsteady turbulent flow through numerical computation, specifically by approximating truncated or filtered solutions to the full incompressible Navier-Stokes equations (\cite{geurts2013direct, Domaradzki2002}). For the simulation of canonical fluid problems, such as sphere-wakes, there are at least three main simulation strategies available: 

\begin{itemize}
    \item Direct numerical simulations (DNS), where the full Navier-Stokes equations are simulated resolving all the length and time scales involved. This approach is limited to diffusive flows with small $Re$, given the elevated computational cost associated to DNS.
    \item Large eddy simulations (LES), proposed by \cite{LES_org}, where the computational cost is reduced by disregarding the small-scale information from the flow, and modeling the unresolved scales introducing a subgrid-scale (SGS) model
    \item Implicit large eddy simulations (ILES), initially proposed by \cite{Germano86_differentialFilters, BORIS1992199}, where the dissipative effects from unresolved scales (introduced by SGS models in LES) are now generated by the intrinsic numerical dissipation associated to the numerical discretization. This numerical dissipation depends on the discretization scheme, and it is the result of truncation errors and applying numerical techniques---such as spectral filtering---designed to maintain numerical stability in the solution. The reader is referred to \cite{GrinsteinILES2002, diamessis_spedding_domaradzki_2011} for a discussion and a review of applications of spectral filtering for ILES strategies. 
\end{itemize}

In general, the results from any of the approaches above need to be compared against observations, experimental data or other numerical studies to evaluate if they are  accurately reproducing the fluid flow dynamics. As summarized in \cite{diamessis_spedding_domaradzki_2011}, DNS investigations 
(\cite{Gourlay2001,BRUCKER_SARKAR_2010, Redford_Lund_Coleman_2015}), LES (\cite{Dommermuth2002}), and ILES (\cite{diamessis_spedding_domaradzki_2011, ZhouDiamessis2019}), have reproduced the basic phenomenology of flow regimes in stratified sphere wakes, particularly for maximum body-based $Re$ values of $Re \sim O(10^4)$ and $Re \sim (10^5)$. Besides corroborating decay laws of kinetic energy dissipation and mean velocity profile, analysis of the results of numerical simulations have permitted the exploration of stratified wake dynamics associated to larger body-based $Re$ values than the one reproduced in laboratory experiments.

\subsubsection{Some of the open questions and objectives}
Although the basic phenomenology of the stratified turbulent wakes is relatively well understood at a qualitative level, the  detailed processes that describe the transitions between flow regimes are not fully clear, particularly at higher $Re$ (\cite{Redfort2015, spedding_1997}). Similarly, exploring how the dynamics and persistence of the coherent vertical structures (or pancake vortices (\cite{spedding_1997})), and any fine-scale turbulence embedded therein, is modified for larger values of $Re$ will give additional insights on how turbulence evolve spatially and temporarily within the SSR (\cite{ZhouDiamessis2019}). It is also necessary to investigate further the role of mixing, viscous dissipation and internal wave radiation in the kinetic energy decay (\cite{KrisPaper}). 

This study intends to address some of these questions by providing a computational tool that will enable us to reproduce the dynamics of stratified wakes at a even higher $Re$ than the ones simulated by \cite{ZhouDiamessis2019}, as high as $\sim O(10^6)$, where turbulent motions are expected to access the SSR in the intermediate to late time, and the radiation of internal waves is known to play an important role in the wake kinetic energy budget. The wake generated by a sphere, which is towed in uniform stratification, serves as a canonical flow to study the structure and dynamics of the turbulent event. In the context of implicit large-eddy simulations ILES, this work describes a numerical model to study the dynamics of turbulent stratified flows. 

\subsection{Numerical model: overview of the flow solver}
The flow solver presented in this work utilizes a pseudo-spectral method for simulating the three-dimensional Navier-Stokes equations under the Boussinesq approximation. Numerical discretizations based on high-order pseudo-spectral methods have emerged as compelling approaches for simulating turbulent stratified flows. Their high accuracy and exponential convergence enables the efficient reproduction of a broad range of length and time scales and,  in schemes based on elemental discretizations, it allows us to localize resolution in areas of interest \cite{Kopriva}. Existing models include triply periodic discretizations (\cite{deBruynKops_2015}, \cite{Winters200469}), and discretizations for complex boundary conditions and geometries using Chebyshev-collocation methods (\cite{Subich_methods_2013}), and Fourier-based techniques combined with higher-order element-based methods (\cite{DIAMESSIS2005}, \cite{Rivera-Rosario2020} and \cite{DIAMANTOPOULOS2022102065}). 

In this work, the numerical model utilizes a Fourier spectral method in the horizontal direction and a modal spectral element method (SEM) discretization in the vertical or wall-normal direction. Using a modal SEM retains the flexibility of localized flow resolution in the vertical. Moreover, in combination with static condensation, the method results in a large number of small tridiagonal systems that allows us to have a Helmholtz solver algorithm that is as inexpensive as second-order finite difference schemes. 
The assumption of a horizontally periodic domain allowed us to use Fourier methods and existing optimized packages for computing Fourier transforms. In the vertical direction, the selection of polynomial basis functions for the spectral element method permitted exploiting matrices structure for performance of the algebraic solvers.
The computer code was designed to use mostly MPI programming and a hybrid MPI/OpenMP approach. Among the main libraries used are FFTW \cite{fftw} and NetCDF \cite{netcdf}. Note that this work is inspired by the implementation of previous models for simulation of the incompressible Navier–Stokes equations, particularly the high-order continuous Galerkin implementation presented in \cite{DIAMANTOPOULOS2022102065}, and the discontinuous spectral multidomain penalty method approach used in \cite{diamessis_spedding_domaradzki_2011}, \cite{DIAMESSIS2005} and \cite{JorgeEscobar2014}.

\subsection{Paper structure}
The text is structured as follows. In Section \ref{methodology}, the model equations are presented, namely the Navier-Stokes equations under the Boussinesq approximation, followed by a description of the numerical discretization. The latter is divided in two parts: $i)$ the temporal discretization (subsection \ref{temporal_dis}) discussing the details of the implicit-explicit splitting scheme that defines the main components of the solver, or fractional steps, to integrate the field variables in time, and $ii)$ the spatial discretization of the governing equations (subsection \ref{spatial_dis}) with a description of the Fourier-based Galerkin method in the horizontal direction, and the modal spectral element method used in the vertical.

Section \ref{benchmark} covers a series of unstratified and stratified Navier-Stokes benchmarks set up to evaluate and quantify the accuracy and stability of the flow solver, including the simulation of a stratified turbulent wake and the comparison of its main features with findings in the literature. Section \ref{sec:conclusions} offers discussion and concluding remarks. Finally, appendices A and B offer additional details on the calculation of nonlinear terms and stabilization techniques.

\section{Equations and discretization} \label{methodology}

\subsection{Mathematical model} \label{mathmodel}

The incompressible Navier-Stokes equations under the Boussinesq approximation (INSEB) are adopted as governing equations, given our interest in studying flows with relatively small density variations. The fluid density is expressed as the sum of three terms: $\rho(x,y,z,t) = \rho_0 + \bar{\rho}(z) + \rho'(x,y,z,t)$, where $\rho_0 \gg \bar{\rho} \gg \rho'$. The quantity $\rho_0$ represents a mean reference value, $\bar{\rho}$ is the horizontally averaged background stratification, and $\rho'$ are the perturbations due to fluid motions \cite{Kundu}. The flow velocity is represented by a vector field $\boldsymbol{u}$, with components $(u,v,w)$ in the $(x,y,z)$ directions, respectively. The INSEB equations are commonly written as,
\begin{align}
\frac{\partial \boldsymbol{u}}{\partial t} + (\boldsymbol{u} \cdot \nabla)\boldsymbol{u} &= - \frac{1}{\rho_0}\nabla p' + \nu \nabla^2 \boldsymbol{u} - \frac{\rho'}{\rho_0} \boldsymbol{g},    \label{INSE_u}\\
\frac{\partial \rho' }{\partial t} + \boldsymbol{u} \cdot \nabla \big ( \rho' + \bar{\rho} \big ) &=   \kappa \nabla^2 \rho', \label{INSE_rho}\\
\nabla \cdot \boldsymbol{u} &= 0, \label{INSE_div}
\end{align}
where $p'$ is the perturbation deviation from the hydrostatic pressure, after the hydrostatic balance has been subtracted from the $z$-momentum equation, and $\nu$ and $\kappa$ are the kinematic viscosity and mass diffusivities. In the density equation, Eq.~(\ref{INSE_rho}), the diffusion of the background density profile $\bar{\rho}(z)$ is ignored in the absence of any external motions.

\subsection{Numerical discretization} \label{numdiscrete}
The temporal discretization of the INSEB equations consists of a high-order fractional-step method. Its spatial discretization is based on the Galerkin principle; the Fourier-Galerkin method is utilized in the horizontal plane or periodic $x$ and $y$ directions, and a Legendre-Galerkin spectral element method in modal form is used in the vertical or non-periodic $z$ direction. 

\subsubsection{Temporal discretization} \label{temporal_dis}
The Karniadakis et al. \cite{Karniadakis1991} scheme, an operator splitting referred to here as KIO, is a variant of the Chorin-Temam method \cite{Jorgethesis}. It consists of fractional steps with a mix of explicit-implicit schemes, specifically Adams-Bashforth methods and backward differentiation formulas \cite{SpectralMethodsIncompressibleViscousFlow}, and high order pressure boundary conditions. Following the notation used in \cite{DIAMANTOPOULOS2022102065}, the result of applying the KIO scheme to Eq.~(\ref{INSE_u}-\ref{INSE_div}), over one time step $\Delta t$, is presented in the form of five steps that allow one to obtain the velocity field $\boldsymbol{u}$, and the scalar field of the density perturbations $\rho'$ at the discrete time $t^{(n+1)}=(n+1)\Delta t$. The five steps are as follows:  

\begin{enumerate}
    \item Computation of nonlinear terms:
        \begin{align}
           \boldsymbol{u}^{*}- \sum_{m=0}^{J_{e}-1} \alpha_m\boldsymbol{u}^{n-m} &= -\Delta t \sum_{m=0}^{J_{\mathrm e}-1} \beta_m \boldsymbol{N} \big(\boldsymbol{u}^{n-m},(\rho')^{n-m} \big ),  \label{eqadvu} \\
           \boldsymbol{N} \big( \boldsymbol{u},\rho' \big ) &=  \boldsymbol{u} \cdot \nabla \boldsymbol{u} + \frac{\rho'}{\rho_0} \boldsymbol{g}, \nonumber
         \end{align}
    where $\boldsymbol{u}^*$ represents an intermediate velocity field with components $(u^*,v^*,w^*)$, and $\alpha=[3,-3/2,1/3]$ and $\beta=[3,-3,1]$ are the time-stepping coefficients of the third order $(J_{\mathrm e}=3)$ KIO scheme \cite{Karniadakis1991}. The operator $\boldsymbol{N} \big( \boldsymbol{u},\rho' \big )$ refers to the sum of the nonlinear terms, written in convective form, and the buoyancy term. Details regarding the spatial discretization and implementation of the non-linear term are provided in \cite{Nidiathesis}. 
    
    To initialize the method, the first time step uses $\alpha=[1,0,0]$ and $\beta=[1,0,0]$, corresponding to time-stepping coefficients of a first order scheme $(J_{\mathrm e}=1)$. The second time step uses coefficient values of $\alpha=[2,-1/2,0]$ and $\beta=[2,-1,0]$, corresponding to a second order scheme $(J_{\mathrm e}=2)$.

    \item Solution of a Poisson equation for the pressure and velocity update:
         \begin{equation}
            \frac{\boldsymbol{u}^{**}- \boldsymbol{u}^{*}}{\Delta t} = - \nabla \check{p}^{n+1/2},  \label{KIOstar}
         \end{equation}
    where $\boldsymbol{u^{**}}$ corresponds to a second intermediate velocity field with components  \\$(u^{**},v^{**},w^{**})$, and $\check{p}^{n+1/2}$ to a time-integrated value of the pressure perturbation averaged over one time-step. By enforcing the incompressibility constraint $\nabla \cdot \boldsymbol{u^{**}}=0$ in Eq.~(\ref{KIOstar}), a Poisson equation is obtained,
        \begin{equation}
             \nabla^2 \check{p}^{n+1/2} = \frac{1}{\Delta t}\nabla \cdot \boldsymbol{u}^{*},
        \label{EqPoisson}   
        \end{equation}
    with high-order pressure boundary conditions as specified in \cite{Karniadakis1991}. Such boundary conditions are aimed towards ensuring high accuracy of the time-discretization scheme, and defined as follows, 
    \begin{align}
    \frac{\partial \check{p}^{n+1/2}}{\partial z} \Bigg |_{z= 0,~L_z}= \boldsymbol{\hat{\mathrm k}} \cdot \Big [ \sum_{m=0}^{J_{\mathrm e}-1} \beta_m \boldsymbol{N} \big(\boldsymbol{u}^{n-m},(\rho')^{n-m} \big )+
    \nu \sum_{m=0}^{J_{\mathrm e}-1} \beta_m (-\nabla \times (\nabla \times \boldsymbol{u})^{n-m}) \big )]. \label{BCPressure}
    \end{align}
    
    Consequently, the solution of $\check{p}^{n+1/2}$ allows one to obtain the vector field $\boldsymbol{u}^{**}$ by reverting to Eq.~(\ref{KIOstar}), and computing, 
     \begin{align}
             \boldsymbol{u}^{**} = \boldsymbol{u}^{*} - \Delta t \nabla \check{p}^{n+1/2}.  
     \end{align}

    \item Viscous term treatment:
        \begin{align}
            \frac{\gamma_0}{\Delta t}~\boldsymbol{u}^{n+1}- \nu~\nabla^2 \boldsymbol{u}^{n+1} &= \frac{1}{\Delta t} \boldsymbol{u}^{**},
            \label{Eq:helmholtzu}
       \end{align}
    where $\gamma_0=11/6$ is a time-stepping coefficient of the third order $(J_{\mathrm e}=3)$ KIO scheme.
    To initialize the method,  the first time step uses $\gamma_0=1$, corresponding to a first order scheme $(J_{\mathrm e}=1)$, and the second time step uses $\gamma_0=3/2$, corresponding to a second order scheme $(J_{\mathrm e}=2)$, as specified in \cite{Karniadakis1991}. Note that the boundary conditions for the velocity field $\boldsymbol{u}^{n+1}$ are enforced in Eq.~(\ref{Eq:helmholtzu}).
 
    \item Solution of advection problem for density perturbations:
        \begin{align}
           \rho^{*}- \sum_{m=0}^{J_{e}-1} \alpha_m (\rho')^{n-m} &= -\Delta t \sum_{m=0}^{J_{\mathrm e}-1} \beta_mR\Big( \boldsymbol{u}^{n-m}, (\rho')^{n-m},\bar{\rho} \Big), \label{eqadvrho} \\ 
            R\big( \boldsymbol{u},\rho', \bar{\rho} \big ) &= \boldsymbol{u} \cdot \nabla \big ( \rho' + \bar{\rho} \big ), \nonumber
        \end{align}
    where $\rho^{*}$ is an intermediate scalar field and $R\big( \boldsymbol{u},\rho', \bar{\rho} \big )$ represents the advective term of the density field. 
    
    \item Solution of a Helmholtz equation for $(\rho')^{n+1}$: 
        \begin{equation}
            \frac{\gamma_0}{\Delta t} (\rho')^{n+1}- \kappa \nabla^2 (\rho')^{n+1} = \frac{1}{\Delta t}\rho^{*}.
            \label{Eq:helmholtzrho}
        \end{equation}
\end{enumerate}
In summary, at every time-step, using the KIO scheme requires the solution of four Helmholtz equations: three for the velocity fields, Eq.~(\ref{Eq:helmholtzu}), and one for the density perturbations, Eq.~(\ref{Eq:helmholtzrho}), one Poisson equation for the mean pressure perturbation, Eq.~(\ref{EqPoisson}), and the computation of four nonlinear advective terms: one for the density perturbations, Eq.~(\ref{eqadvrho}), and three for the velocity fields, Eq.~(\ref{eqadvu}).

\subsubsection{Spatial discretization} \label{spatial_dis}

\paragraph{Discretization in the periodic directions: Fourier-Galerkin method.}

The periodic directions $x \in [0,L_x]$ and $y \in [0,L_y]$ define a two-dimensional domain denoted as $\Omega' \subset \mathbb{R}^2$. $\hat{\Omega} \subset \mathbb{R}^3$ denotes the extruded three-dimensional domain in the non-periodic direction $\hat{\Omega}=\Omega' \times [0,L_z]$. As an example, the discretization of the first velocity component $u(x,y,z)$ will be presented below, as the other velocity components and field variables $\check{p}$ and $\rho'$ are treated in the same way.

The solution $u(x,y,z)$ at a given time-step is assumed to be approximated by a truncated Fourier series in the $x$ and $y$ directions,
\begin{equation}
    u = u(x,y,z) \simeq \sum_{j=-N_y/2}^{N_y/2}\sum_{i=-N_x/2}^{N_x/2} \tilde{u}(\mathrm k_x, \mathrm k_y, z) \phi(\mathrm k_x, \mathrm k_y, x, y),
    \label{fourierExp}
\end{equation}
where $\phi(\mathrm k_x, \mathrm k_y, x,y)=e^{ \mathrm{\mathrm{i}\mkern1mu} ( \mathrm k_x x + \mathrm k_y y)}$ are the Fourier basis functions, and $\mathrm k_x=(2\pi/L_x) i$ and $\mathrm k_y=(2\pi/L_y) j$ are the wavenumbers in the $x$ and $y$ directions, respectively. The unknown Fourier coefficients are represented by $\tilde{u}(\mathrm k_x,\mathrm k_y,z) \in \mathbb{C}$, and $N_x$ and $N_y$ are the number of grid points in $x$ and $y$, respectively. Note that $\tilde{u}(\mathrm k_x, \mathrm k_y, z)$ has complex-conjugate symmetry about $(k_x, k_y, z) = (0, 0, z)$ due to the fact that $u(x, y, z) \in \mathbb{R}$. Therefore, at a given $z$, one half-plane of the Fourier coefficients determines the rest, so that (for example) only $N_y/2$ Fourier coefficients in $y$, and $N_x$ Fourier coefficients in $x$, are required to represent $u$ in Eq.~(\ref{fourierExp}). In software, we make use of this more compact representation by choosing the first Fourier approximation to be computed in the $y$ direction (real-valued Fourier transformation, in which complex conjugation is implied).

Using a Fourier-Galerkin approximation \cite{Kopriva}, and by virtue
of the orthogonality of the Fourier basis functions, the first Helmholtz problem of Eq.~(\ref{Eq:helmholtzu}) is written in semi-discrete form below. Note that the wavenumber dependence of $\tilde{u}^{n+1}$ and right-hand-side $\tilde{u}^{**}$ are not included here to simplify notation. For a single horizontal wavenumber pair $(\mathrm k_x,\mathrm k_y)$, we write the $x$ component of Eq.~(\ref{Eq:helmholtzu}) as, 

\begin{equation}
    -\nu \frac{\partial ^2}{ \partial z^2 } \tilde{u}^{n+1} + \big( \nu |\mathbf{k}|^2 + \frac{\gamma_0}{\Delta t} \big) \tilde{u}^{n+1} = \frac{1}{\Delta t} \tilde{u}^{**},
    \label{firstHelmholtz}
\end{equation}
where $\tilde{u}^{n+1}=\tilde{u}^{n+1}(\mathrm k_x,\mathrm k_y,z) \in \mathbb{C}$, and $\tilde{u}^{**}=\tilde{u}^{**}(\mathrm k_x,\mathrm k_y,z) \in \mathbb{C}$ are the $(\mathrm k_x,\mathrm k_y)$ wavenumber pair Fourier coefficients representation of $u^{n+1}$ and $u^{**}$, respectively. The z contribution of the Laplacian operator is represented by ${\partial^2}/{ \partial z^2 }$, and its $x$ and $y$ components are written now in terms of the magnitude of the wavenumber pair $|\mathbf{k}|^2=(\mathrm k_x^2+\mathrm k_y^2)$. Such discretization now represents a series of Helmholtz problems in Fourier space that are independent of each other: one Helmholtz problem for every $(\mathrm k_x,\mathrm k_y)$ wavenumber pair. A similar approach is followed to write the discrete form of the second and third components of Eq.~(\ref{Eq:helmholtzu}), as well as the Helmholtz problems for $\check{p}^{n+1/2}$, Eq.~(\ref{EqPoisson}), and $(\rho')^{n+1}$, Eq.~(\ref{Eq:helmholtzrho}). 

The nonlinear term $\boldsymbol{N} \big( \boldsymbol{u},\rho' \big )$ in Eq.~(\ref{eqadvu}) is computed in a pseudo-spectral manner, as detailed in section 3.2.1. of \cite{canuto1988}. The gradient $\nabla \boldsymbol{u}$ is calculated in spectral space (Fourier-Lobatto space), but the multiplication $\boldsymbol{u}\cdot \nabla \boldsymbol{u}$ (expressed as a convective flux) is performed pointwise in physical space using discrete transforms. Stabilization techniques are discussed in Appendix A. Additional details on the treatment of the nonlinear terms can be found in \cite{Nidiathesis}. 

\paragraph{Discretization in the non-periodic direction: Legendre-Galerkin spectral element method.}
A continuous Galerkin finite element method is adopted, where additional degrees of freedom are added to the elements, and grid nodes are shared at the element interfaces. The local polynomial representation, however, does not correspond to a nodal expansion---where the interpolating Lagrange polynomial is used to represent the solution---but to a modal one, where a boundary-adapted basis \cite{canuto2007spectral} is used instead. For more details regarding boundary-adapted bases and the differences between nodal and modal expansions, the reader is referred to \cite{canuto2007spectral}.
The non-periodic domain defined by $z \in [0,L_z]$, $\Omega \subset \mathbb{R}$, is represented by an assembly of $N_e$ elements that populate the domain in the vertical direction. Let $\mathcal{V} \subset H^1 $ be a finite subspace spanned by polynomial functions up to order $N_p-1$, $\mathcal{V}=\textrm{span}\{\Psi_0(z),\ldots,\Psi_{N_{\text{dof}}}(z)\}$. 
The solution $\tilde{u}=\tilde{u}(\mathrm k_x,\mathrm k_y,z)$ from Eq.~(\ref{fourierExp}), is approximated now by a sum of piecewise basis or expansion functions $\Psi_i(z)$: $\tilde{u}=\sum_{i} \hat{u}_{i}{\Psi_i(z)}$, where $i \in \{1,\hdots,N_{\text{dof}}\}$ is the index set, $\boldsymbol{\hat{u}}=\{ \hat{u}_1,...,\hat{u}_{N_{\text{dof}}} \}$ is the modal representation of $\tilde{u}$, and $N_{\text{dof}}$ is the total (global) number of degrees of freedom in $\Omega$. The global functions $\Psi_i(z)$ are expressed in terms of local functions $\psi_p(z)$ in every element,
\begin{equation}
\tilde{u}=\sum_{i=1}^{N_{\text{dof}}} \hat{u}_{i}{\Psi_i(z)}=\sum_{e=1}^{N_e} \sum_{p=0}^{N_p-1} \hat{u}_{p}^e{\psi_p^e(z)},
\label{EqallExpansion}
\end{equation}
where $N_p$ is the polynomial order of the expansion, $N_e$ is the number of elements, and the superscript $e$ denotes the element in which the $N_p-1$ order local expansion takes place. Since the method is continuous, every element shares one mode with each of its neighbouring elements (i.e. $\hat{u}_{N_{p}-1}^1=\hat{u}_{0}^2$ and $\hat{u}_{N_{p}-1}^2=\hat{u}_{0}^3$). Note that the global number of degrees of freedom $N_{\text{dof}}=N_e\times (N_p-1)+1$ is smaller than than the total number of local degrees of freedom $M_z=N_e\times N_p$. A procedure known as direct stiffness summation, or global assembly (see \cite{karniadakisSpectralMethods,deville_fischer_2002}), reassembles the global expansion from the local expansion on every element. Then, operations can be performed locally inside the elements and assembled to compute global operations---such as integrals---in $\Omega$. The reader is referred to \cite{karniadakisSpectralMethods} for additional details on the fundamental concepts behind finite element methods and direct stiffness summation.

\begin{figure}
    \centering
    \includegraphics[scale=0.5]{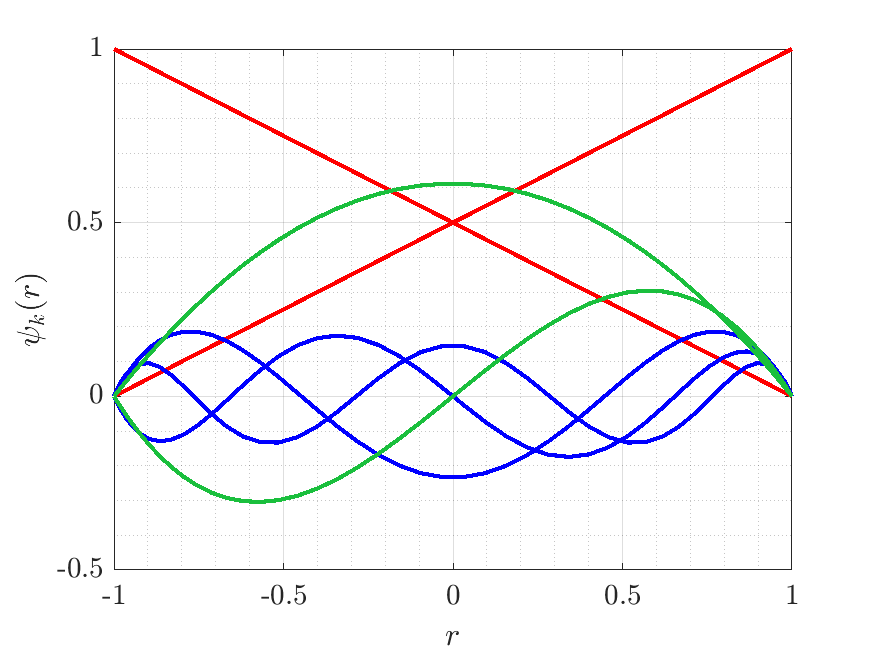} 
   \caption{First seven modes of the modal boundary-adapted basis described by Eq.(\ref{Eq:lobattoBasis}). Linear or vertex modes $\psi_0(r)$ and $\psi_1(r)$ are colored in red. First two bubble modes $\psi_2(r)$ and $\psi_3(r)$ are colored in green, and higher order bubble modes $\psi_k(r)$, $4\leq k \leq 6$ in blue. Color code is introduced to facilitate the interpretation of the matrix sparsity patterns in Fig.~\ref{sparsityPatterns}.}
   \label{figLobatto}
\end{figure}

In the reference domain $r \in [-1,1]$, $L_k$ is the Legendre polynomial of $k^{th}$ order that satisfies the recursion relation, 
\begin{align}
 L_0(r)&= 1 \nonumber \\
 L_1(r)&= r \nonumber \\
 L_{k+1}(r)&= \frac{2k+1}{k+1}rL_{k}(r) - \frac{k}{k+1}L_{k-1}(r), & 2\leq k \leq N. 
\label{Eq:legendreBasis}
\end{align}
Then, the boundary-adapted basis of modal type $\psi_k$ (\cite{canuto2007spectral}) is defined as, 
\begin{align}
 \psi_0(r)&= \frac{1}{2}(L_{0}(r)-L_{1}(r)) \nonumber \\
 \psi_1(r)&= \frac{1}{2}(L_{0}(r)+L_{1}(r)) \nonumber \\
 \psi_k(r)&= \frac{1}{\sqrt{2(2k-1)}}(L_{k-2}(r)-L_{k}(r)), & 2\leq k \leq N. 
\label{Eq:lobattoBasis}
\end{align}
The basis $\psi_k$, described in detail in \cite{canuto2006}, was originally proposed by Babu\u{s}ka and collaborators \cite{edsjsr.215686919810601} to be used in the p-version of the finite-element method. Non-orthogonal by design, it is composed of two linear functions (or boundary basis functions $\psi_0$ and $\psi_1$) that are nonzero at the end of the interval, and a series of internal or bubble functions that oscillate in the interior and vanish at both endpoints. The first seven modes of $\psi(r)$ are illustrated in Fig.~\ref{figLobatto}. An affine mapping, $z(r)$, is introduced to write the local modal expansion from Eq.~(\ref{EqallExpansion}), $\sum_{p} \hat{u}_{p}^e{\psi_p^e \big( z(r) \big )}$, in every element. The mapping $z(r)= z_l^e+\frac{1+r}{2} h^e$, where $h^e=z_r^k-z_l^k$ is the element length and $z_r^e$ and $z_l^e$ are the element interfaces, supports elements that non-uniformly spaced.  

Under the Galerkin approximation, where the test functions $\mathrm{\upsilon}$ are chosen to be $\upsilon=\sum_{i} \hat{\upsilon_i} \Psi_i(z)$---with $i \in \{1,\hdots,N_{\text{dof}}\}$, the weak form of the Helmholtz problem in Eq.~(\ref{firstHelmholtz}) is written as follows,
\begin{equation}
a \int_{0}^{L_z} \frac{\partial \upsilon}{\partial z} \frac{ \partial \tilde{u}^{n+1}}{\partial z}~dz + b \int_{0}^{L_z} \upsilon \tilde{u}^{n+1}~dz = \int_{0}^{L_z} \upsilon \tilde{u}^{**}~dz + a\Big [ \upsilon \frac{\partial \tilde{u}^{n+1}}{\partial z} \Big ]_0^{L_z}, \label{intByParts}
\end{equation}
where $a=-\nu$ and $b=( \nu|\mathbf{k}|^2 + \gamma_0/\Delta t \big )$. Assuming homogeneous Neumann boundary conditions for $\tilde{u}$, the matrix form of Eq.~(\ref{intByParts}) is expressed as,
\begin{equation}
\Big(a K+ b M  \Big) \boldsymbol{\widehat{u}}^{n+1} = \boldsymbol{\widehat{f}} \Leftrightarrow 
A \boldsymbol{\widehat{u}}^{n+1} = \boldsymbol{\widehat{f}}
\label{discHelm}
\end{equation}
where  $\boldsymbol{\widehat{u}}^{n+1}=\boldsymbol{\widehat{u}}^{n+1}(\mathrm k_x,\mathrm k_y) \in \mathbb{C}$ is the vector of modal coefficients of $\tilde{u}^{n+1}=\tilde{u}^{n+1}(\mathrm k_x,\mathrm k_y,z)$, $\boldsymbol{\widehat{f}}=\boldsymbol{\widehat{f}}(\mathrm k_x,\mathrm k_y)\subset \mathbb{C}$ is the right hand side vector $\boldsymbol{\widehat{f}}=M \boldsymbol{\widehat{u}}^{**}+\boldsymbol{g}$ , and $A=(a K + b M)$ is the resulting Helmholtz operator. Note that $K$ and $M$ correspond to the global assembled stiffness and mass matrices (\cite{deville_fischer_2002}) representing the inner products,
\begin{align}
K_{ij}&=\int_{\Omega} \frac{d\Psi_i}{dz} \frac{d\Psi_j}{dz}dz, \\
M_{ij}&=\int_{\Omega} \Psi_i \Psi_j dz, \label{globalMassMatrix}
\end{align}
where $i,j \in \{1,...,N_{\text{dof}}\}$. The entries of the matrices $K$ and $M$ that correspond to each element (local matrices) can be computed exactly, following the expressions presented in section 8.5 of \cite{canuto2007spectral}, with a subsequent assembly of the shared element interfaces, using direct stiffness summation. In regard to the right-hand side vector, an $L^2$ projection on the basis functions $\psi(z)$ is used to obtain the right-hand-side coefficients $\boldsymbol{\widehat{u}}^{**}=\{ \widehat{u}^{**}_1,...,\widehat{u}^{**}_{N_{\text{dof}}} \}$  from the known values of $\tilde{u}^{**}=\tilde{u}^{**}(\mathrm k_x,\mathrm k_y,z)$, where $\tilde{u}^{**}=\sum_{i} \hat{u}^{**}_{i}{\Psi_i(z)}$.

Dirichlet boundary conditions in Eq.~(\ref{discHelm}) are enforced by updating the right-hand side vector in a procedure known as ``lifting" (\cite{karniadakisSpectralMethods}), while Neumann boundary conditions are included in the weak formulation naturally through integration by parts (see Eq.~(\ref{intByParts})). The contribution of the Dirichlet boundary conditions prescribed for $u^{n+1}$ in Eq.~(\ref{discHelm}) is incorporated in the term $\boldsymbol{g}$ on the right-hand side. For details regarding implementation of Dirichlet boundary conditions in the weak form, the reader is referred to \cite{karniadakisSpectralMethods}. 

A similar weak-form-based formulation is used to write the discrete form of the Helmholtz problems for the $v^{n+1}$ and $w^{n+1}$ velocity components in Eq.~(\ref{Eq:helmholtzu}), as well as for $\check{p}^{n+1/2}$, Eq.~(\ref{EqPoisson}), and $(\rho')^{n+1}$, Eq.~(\ref{Eq:helmholtzrho}). Note that the values of $a$ and $b$ of every Helmholtz operator $A=(a K + b M)$ are varied according to each Helmholtz problem. For example, for every $(\mathrm k_x,\mathrm k_y)$ wavenumber pair, the discrete form of the pressure Eq.~(\ref{EqPoisson}) is written as,
\begin{align}
\Big( a{K}+b{M} \Big) \boldsymbol{\widehat{p}}^{n+1/2}&=\boldsymbol{  \widehat{f_p}}, \label{eqHelmholtzp2}
\end{align}
\noindent where $\boldsymbol{\widehat{p}}^{n+1/2}=\boldsymbol{\widehat{p}}^{n+1/2}(\mathrm k_x,\mathrm k_y) \in \mathbb{C}$ is the modal representation of $\check{p}^{n+1/2}$, and $\boldsymbol{\widehat{f_p}}=\boldsymbol{\widehat{f_p}}(\mathrm k_x,\mathrm k_y) \in \mathbb{C}$ is the right hand side $\boldsymbol{\widehat{f_p}}=\frac{1}{\Delta t}M (\nabla \cdot \boldsymbol{\widehat{u}}^{*})+\boldsymbol{q}$. The term $\boldsymbol{q}$ accounts for the contribution of the Neumann boundary conditions for the pressure as defined in Eq.~(\ref{BCPressure}). In this case, the Helmholtz operator $A=(a K + b M)$ utilizes $a=1$ and $b=|\mathbf{k}|^2$.

Similarly, the discrete form of the density Eq.~(\ref{Eq:helmholtzrho}),
\begin{align} 
\Big(a{K}+ b{M}  \Big) \boldsymbol{\widehat{\rho}}^{n+1}&= \boldsymbol{\widehat{f_{\rho}}},\label{helmholtzRho}
\end{align}

where $\boldsymbol{\widehat{\rho}}^{n+1}=\boldsymbol{\widehat{\rho}}^{n+1}(\mathrm k_x,\mathrm k_y) \in \mathbb{C}$ is the modal representation of $(\rho')^{n+1}$, and the right-hand-side $\boldsymbol{\widehat{f_\rho}}(\mathrm k_x,\mathrm k_y)$ is computed as $\boldsymbol{\widehat{f_\rho}}=\frac{1}{\Delta t}M (\nabla \cdot \boldsymbol{\widehat{u}}^{*})$. Throughout this work, homogeneous Neumann boundary conditions are prescribed for $(\rho')^{n+1}$. The Helmholtz operator $A=(a K + b M)$ in Eq.~(\ref{helmholtzRho}) is defined by $a=\kappa$ and $b=(\kappa |\mathbf{k}|^2 + \gamma_0/\Delta t)$.

One of the main advantages of the selection of particular basis functions, $\psi(z)$, is the emergence of global mass and stiffness matrices that are banded: ${K}$ is triadiagonal but can be alternatively represented in terms of a diagonal and a tridiagonal block (or submatrix), and ${M}$ has a pentadiagonal structure \cite{canuto2007spectral} that can be described in terms of two tridiagonal blocks and coupling coefficients. Sparsity patterns of the stiffness and mass matrices are illustrated in Fig.~\ref{sparsityPatterns} for conventional (lexicographic) and alternative ordering/indexing of unknowns. The alternative ordering results in an favorable banded matrix structure that will be exploited for performance in the implementation of the Helmholtz solver (see Sec. \ref{sec:helmholtz}).  

\subsection{Helmholtz solver: domain decomposition and Schur complement in the non-periodic domain} \label{sec:helmholtz} \mbox{} 
The solution of the linear systems in the form of Eq.(\ref{discHelm}) (such as Eq.~(\ref{eqHelmholtzp2}) or (\ref{helmholtzRho})) leverages two techniques: Gaussian elimination and static condensation, also known as substructuring \cite{deville_fischer_2002}, where the unknowns are ordered in groups based on their location in the computational domain $\Omega$. The unknowns are grouped by collecting first the interior or bubble modes and then the vertex or boundary modes. The discrete Helmholtz operator $A=(aK+bM)$ is represented as a block matrix,

\begin{equation}
\begin{bmatrix}
   \color{blue} \boldsymbol{A_{II}}  & \color{ForestGreen} \boldsymbol{A_{IB}}  \\
    \color{ForestGreen} \boldsymbol{A_{BI}}  &  \color{red} \boldsymbol{A_{BB}}  \\
\end{bmatrix} 
\begin{bmatrix}
   \widehat{u_I}  \\
   \widehat{u_B}  \\
\end{bmatrix} =
\begin{bmatrix}
   \widehat{f_I}  \\
   \widehat{f_B}  \\
\end{bmatrix}
\label{basicmatrix}
\end{equation}
where $A_{II}$ contains the interior mode contributions for all the elements, $A_{BB}$ the element boundary modes coefficients, and $A_{IB}$ and $A_{BI}$ represent the coupling between the interior and the boundary modes. In Eq.~(\ref{basicmatrix}), the unknowns $\widehat{u_I}$, $\widehat{u_B}$ and right hand side entries $\widehat{f_I}$, $\widehat{f_B}$ are indexed accordingly.

\begin{figure}
    \centering
    \begin{tabular}{c c}
    \includegraphics[scale=0.46]{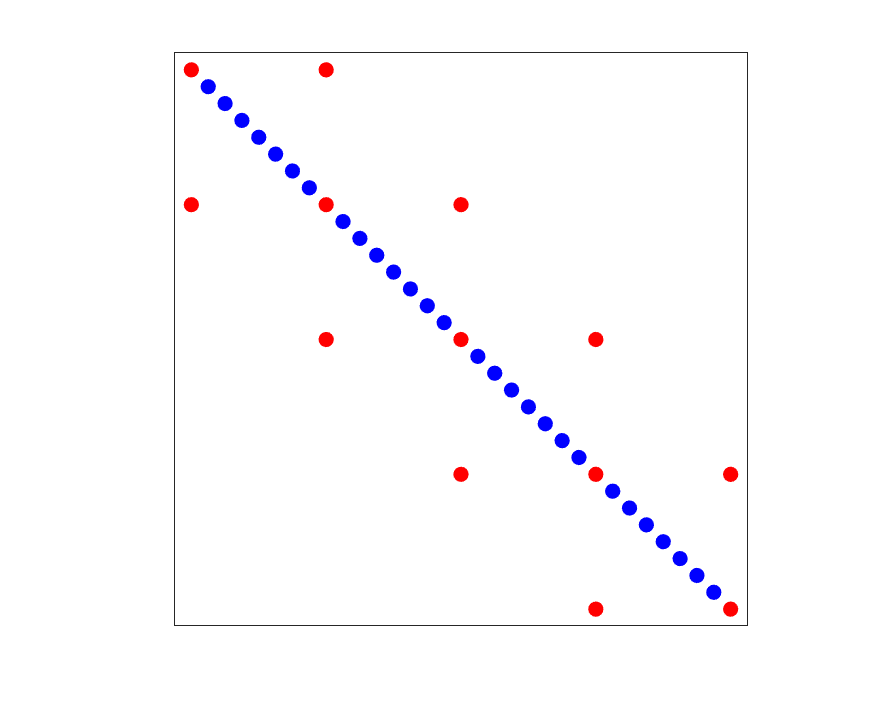} &
    \includegraphics[scale=0.46]{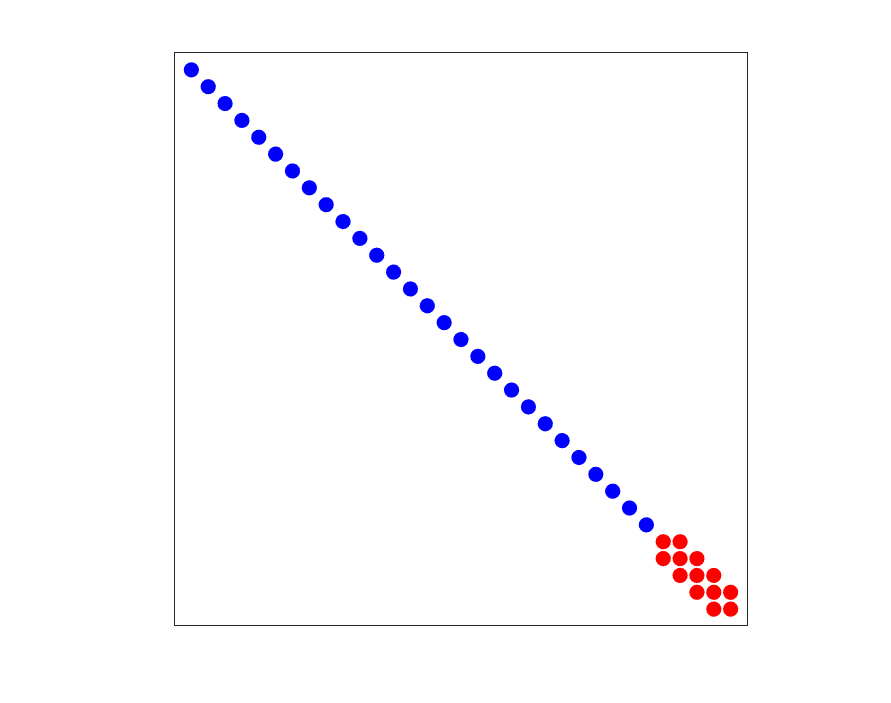} \\
    \includegraphics[scale=0.46]{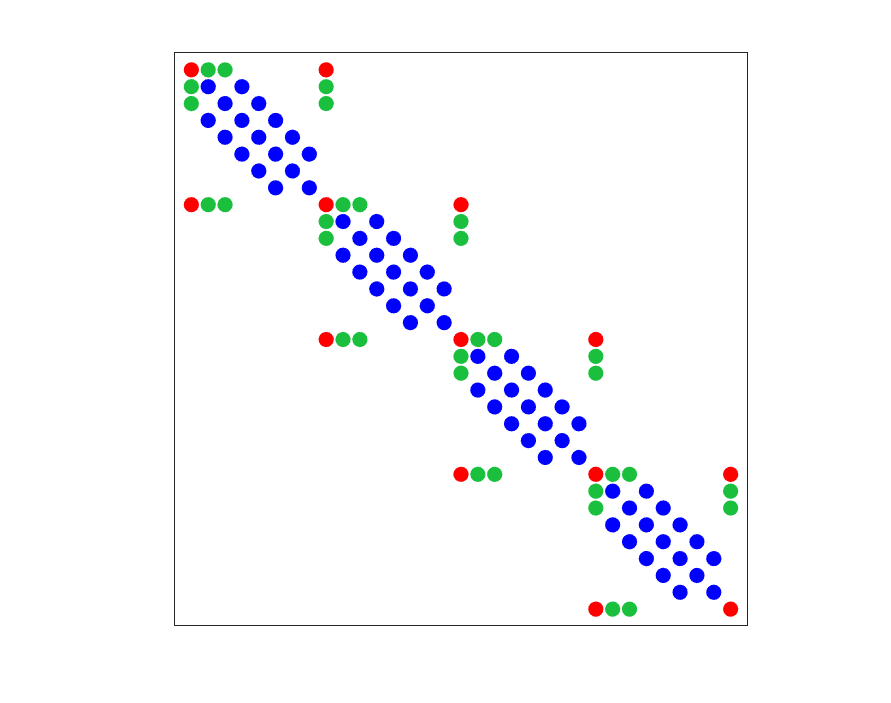} &
    \includegraphics[scale=0.46]{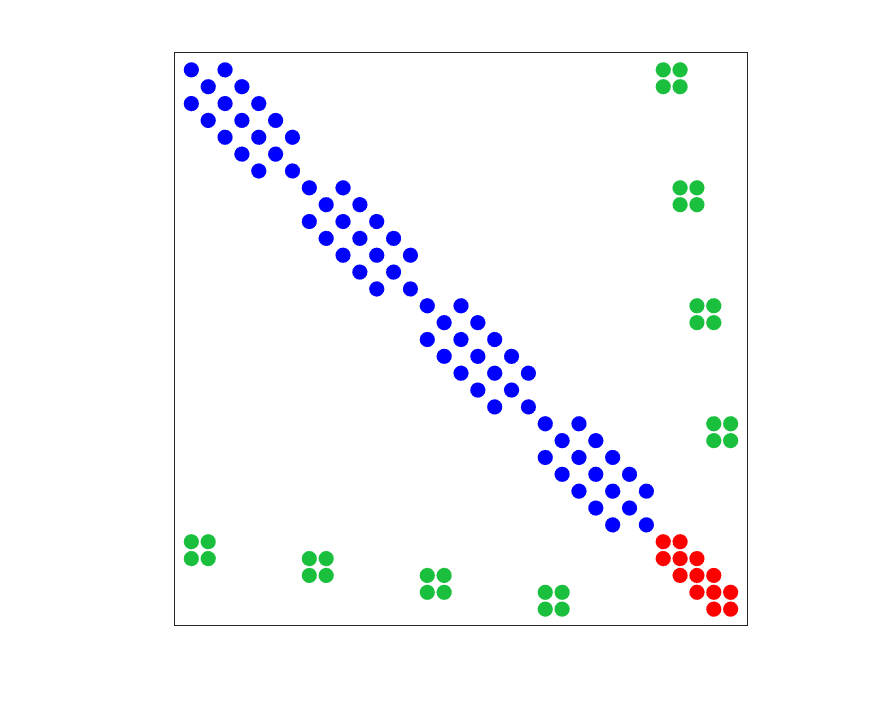} 
   \end{tabular} 
    \begin{tabular}{c c}
     
   \end{tabular} 
  \caption{Sparsity patterns of the global second order stiffness matrix ${K}$ (top) and global mass matrix ${M}$ (bottom) using the local modal boundary-adapted basis defined in Eq.~(\ref{Eq:lobattoBasis}), and four elements with nine modes per element. Left: Lexicographic order. Right: separating interior from boundary modes. Dots indicate terms corresponding to edge modes (red), interior modes (blue), and the coupling terms between them (green).}
   \label{sparsityPatterns}
\end{figure}

The result of grouping the unknowns in this way is the emergence of sparsity patterns that are conducive to increasing the performance of the Helmholtz solver. As can be seen in Fig.~\ref{sparsityPatterns}, where interior, boundaries, and coupling entries have been colored differently and reorganized, the resulting matrix blocks have a tridiagonal structure, which allows us to employ a commonly used form of Gaussian elimination for tridiagonal systems: the Thomas algorithm. 

To solve the linear system of Eq.~(\ref{basicmatrix}),  first, we solve for the coefficients corresponding to the boundaries $\widehat{u_B}$, and second for the interiors $\widehat{u_I}
$. In the first stage of the domain decomposition, the unknowns $\widehat{u_I}$ are eliminated via a block Gaussian elimination, and Eq.(\ref{basicmatrix}) is condensed into a Schur complement system, 
\begin{align}
S^{(1)}\widehat{u_B}=g^{(1)}
\label{SchurSystem1}
\end{align}
where $S^{(1)}$ is the Schur complement matrix and $g^{(1)}$ is its right hand side. $S^{(1)}$ and $g^{(1)}$ are computed as follows,
\begin{align}
S^{(1)} &= \color{red} \boldsymbol{A_{BB}} \color{black} - \color{ForestGreen} \boldsymbol{A_{BI}}\color{blue} \boldsymbol{(A_{II})^{-1}}\color{ForestGreen} \boldsymbol{ A_{IB}}\color{black} \\
g^{(1)} &= \widehat{f_B}- \color{ForestGreen} \boldsymbol{A_{BI}} \color{blue} \boldsymbol{(A_{II})^{-1}}\color{black} \widehat{f_{I}}
\end{align}

Solving the linear system of Eq.~(\ref{SchurSystem1}) requires the inversion of $A_{II}$, which is performed by computing an $LDL^T$ factorization, and using backward and forward substitution. Note that the solution of $\widehat{u_I}$ and $\widehat{u_B}$ only involves tridiagonal systems. The particular tridiagonal structure of $A_{II}$, where the interior modes of each element are decoupled, enables performing the inversion $A_{II}^{-1}$ for all the elements in parallel. In general, the entries of $S^{(1)}$ are computed in an element-wise fashion, with a subsequent assembly of the shared element interfaces using direct stiffness summation \cite{deville_fischer_2002}. 
The discrete Helmholtz operator $A=(aK+bM)$ that results from the Poisson problem in Eq.(\ref{EqPoisson}), where $a=1$ and $b=|\boldsymbol{k}|^2$, involves pure Neumann boundary conditions. When solving for the horizontal wavenumber pair $(k_x,k_y)=(0,0)$, the resulting operator, $A=K$, is singular and requires regularization to fulfill the solvability condition (see \cite{POZRIKIDIS2001} and \cite{POZRIKIDIS1997}). Specifically for this case, an additional level of static condensation is applied over the tridiagonal system $S^{(1)} \widehat{u_B} = g^{(1)}$. The grid points corresponding to element interfaces are organized into two groups,

\begin{equation}
\begin{bmatrix}
   S^{(1)}_{II}  & S^{(1)}_{IE}  \\
   S^{(1)}_{EI}  & S^{(1)}_{EE}  \\
\end{bmatrix} 
\begin{bmatrix}
   \widehat{u}_{B_{I}}  \\
   \widehat{u}_{B_{E}}  \\
\end{bmatrix} =
\begin{bmatrix}
   g^{(1)}_{I}  \\
   g^{(1)}_{E}  \\
\end{bmatrix}
\label{Schur2}
\end{equation}
where unknowns corresponding to the two-element interfaces at the boundaries of the computational domain are identified as $\widehat{u}_{B_E}$, and the element interfaces located in the interior of the domain are identified as $\widehat{u}_{B_I}$. The second Schur complement system corresponds to the Schur complement matrix $S^{(2)}$ and its right-hand side $g^{(2)}$,
\begin{align}
S^{(2)} = S^{(1)}_{EE} - S^{(1)}_{EI} \big ( S^{(1)}_{II} \big )^{(-1)} S^{(1)}_{IE} \\
g^{(2)} = g^{(1)}_{E}-S^{(1)}_{EI} \big ( S^{(1)}_{II} \big )^{(-1)} g_I^{(1)}
\end{align}

 The required normalization is applied to a reduced system $S^{(2)}\widehat{u}_{B_E}=g^{(2)}$ that only contains interactions of the leftmost boundary mode with the rightmost boundary mode. By static condensation, instead of regularizing the original system of the rank of $S^{(1)}$ (total number of elements in $\Omega$), one regularizes a system of rank 2. The treatment of the singularity is done using an augmented (bordered) system \cite{Jorgethesis}, where the null singular value is removed by applying a correction to the right-hand-side vector \cite{POZRIKIDIS2001}. For details regarding the treatment of the singular system, the reader is referred to \cite{POZRIKIDIS1997} and \cite{multigrid}.

\section{Navier-Stokes benchmarks} \label{benchmark}

Results of a series of benchmark problems are used to validate the flow solver. The benchmark problems are listed in increasing complexity, allowing for a progressive evaluation of the solver's capabilities. Accuracy is assessed by comparing global quantities of the flow, such as domain-integrated kinetic energy and enstrophy, against published studies or experimental data. The first benchmark involves a two-dimensional and three-dimensional collision of a vortex dipole with a no-slip wall. This is followed by a more complex two-dimensional case of an internal solitary wave in uniform-depth water.

\subsection{Normal collision of a vortex dipole with a no-slip wall}
The accurate simulation of the collision of a dipole with a no-slip wall, as described by \cite{Clercx06}, has become an appropriate test case to evaluate the reliability of numerical implementations of the INSEB equations. This benchmark tests the solver's capabilities to predict the trajectories of two mutually interacting monopoles and reproduce the formation of finer-scale vortex features in the boundary layer. Using the same approach followed by \cite{DIAMANTOPOULOS2022102065} and \cite{Subich_methods_2013}, a two-dimensional collision is first simulated, followed by an extension to three dimensions. Accuracy is measured by computing global quantities of the flow, such as total kinetic energy, enstrophy, and vorticity distribution at the boundary.

\begin{figure}
\centering
     \subfloat[$t = 0.0$\label{fig:3a}]{\includegraphics[scale=0.225, trim={0cm 0cm 6.5cm 0cm},clip]{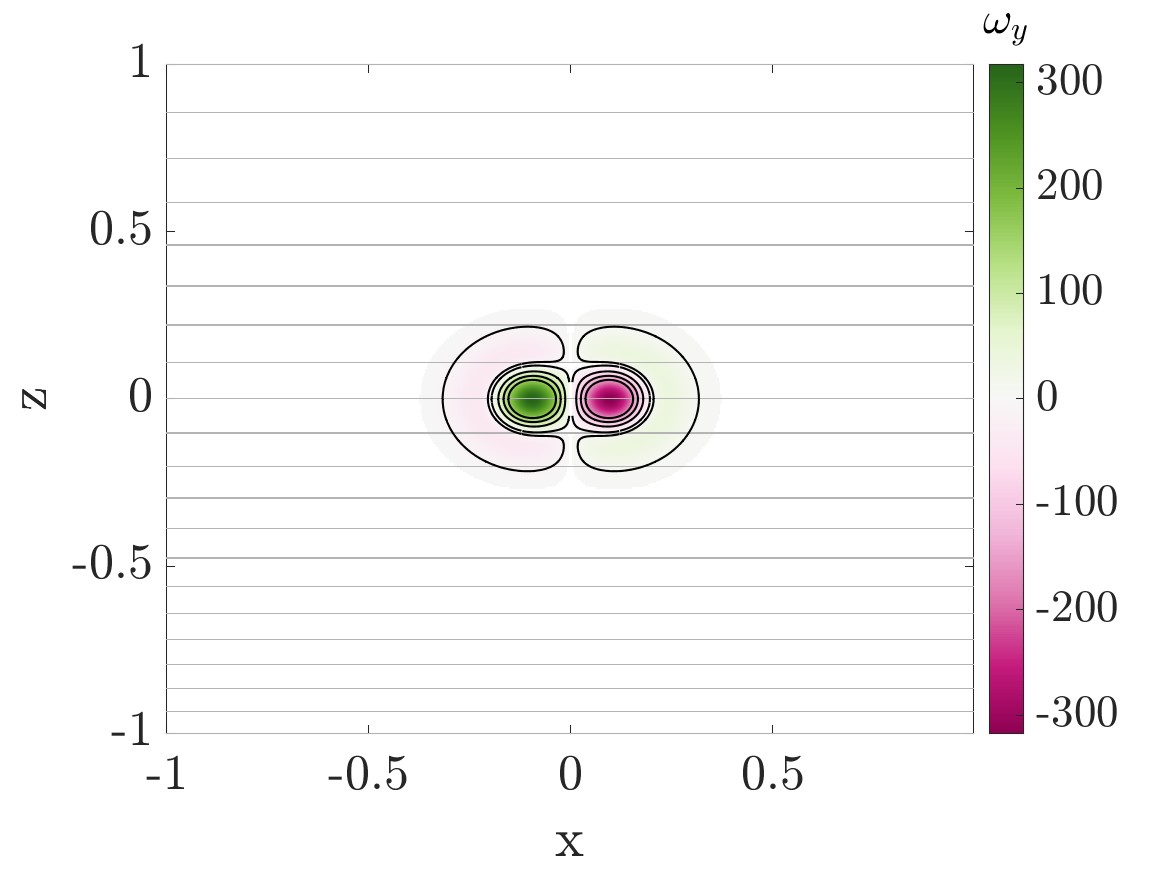}}
    \subfloat[$t = 0.25$\label{fig:3b}]{\includegraphics[scale=0.3, trim={3.8cm 0cm 0cm 0cm},clip]{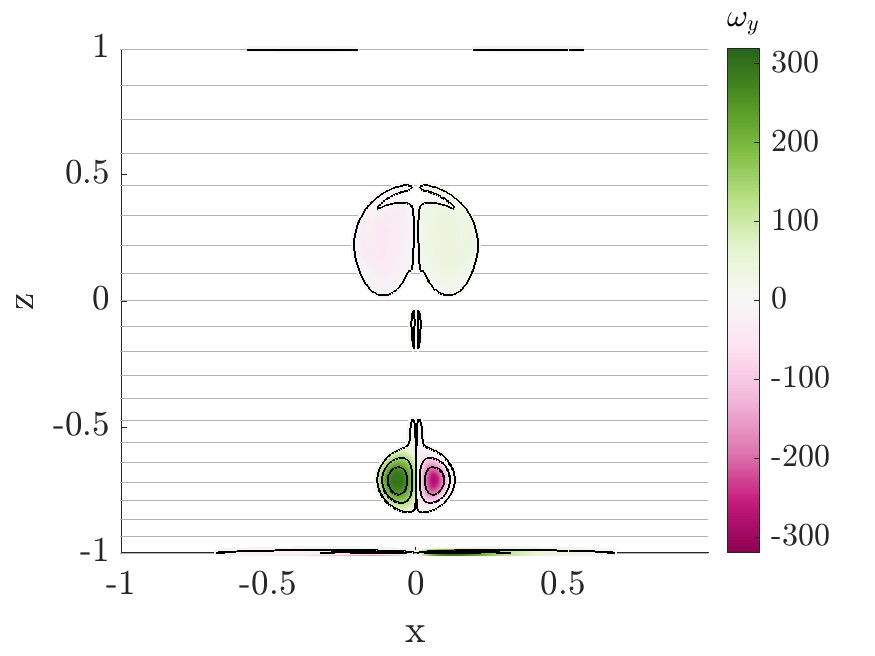}} \\
    \subfloat[$t = 0.36$\label{fig:3c}]{\includegraphics[scale=0.305, trim={0cm 0cm 5.2cm 0cm},clip]{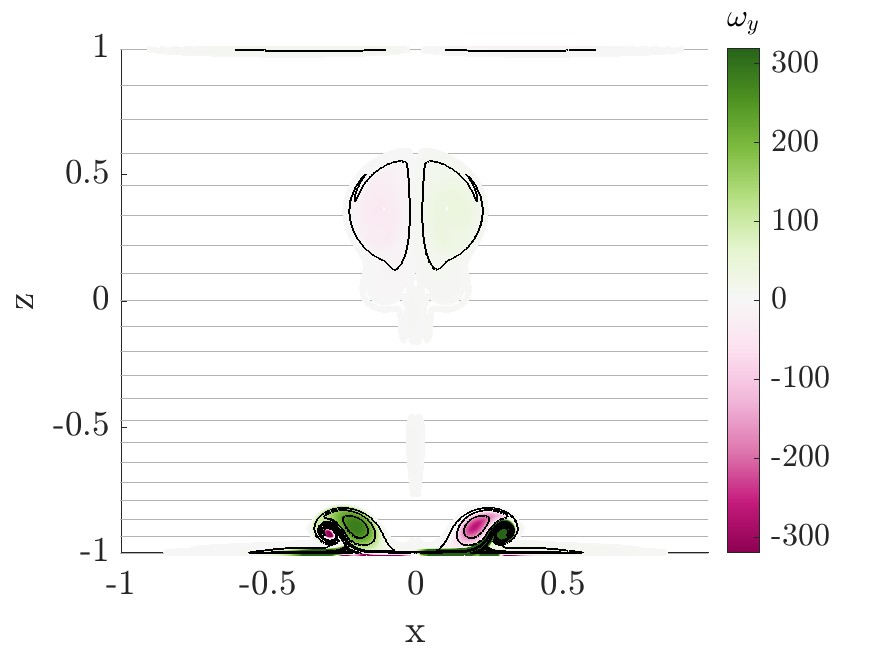}}
    \subfloat[$t = 0.51$\label{fig:3d}]{\includegraphics[scale=0.23, trim={5.3cm 0cm 0cm 0cm},clip]{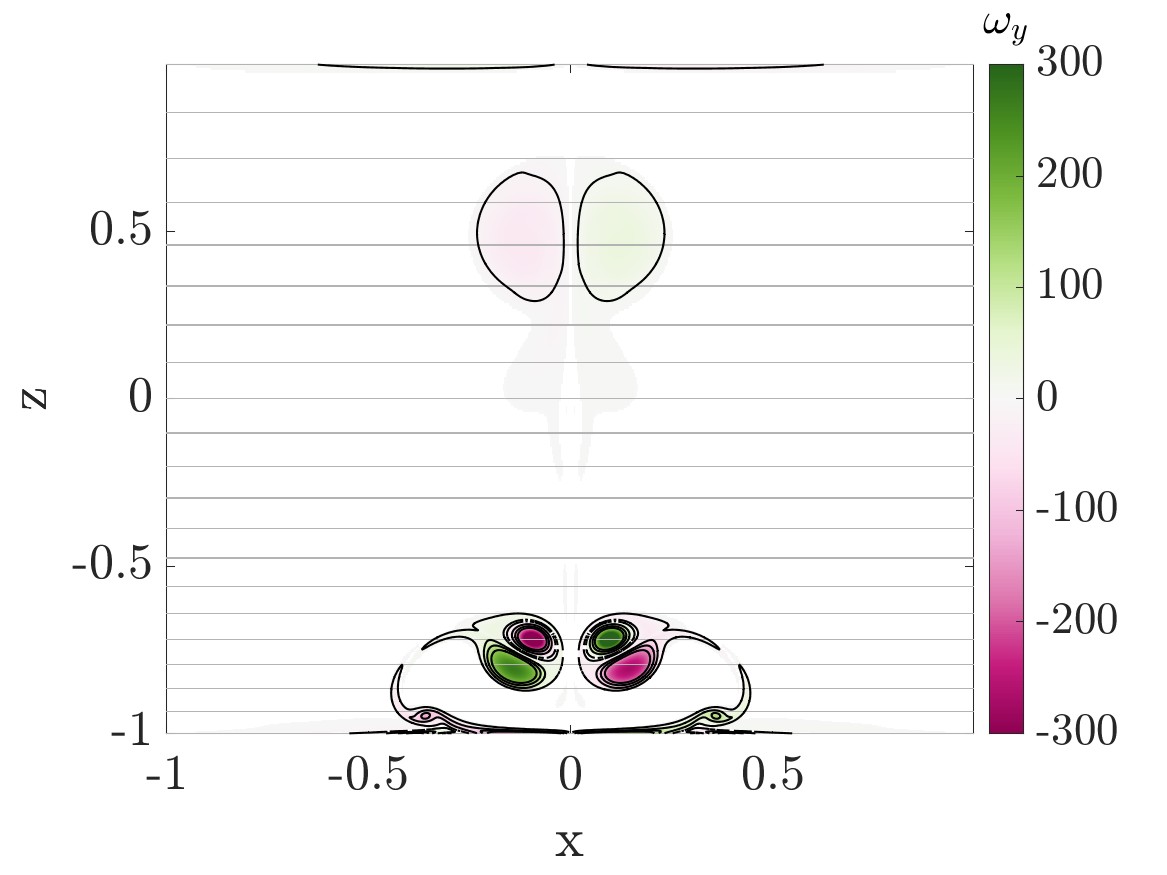}}
    \caption{Vorticity field of translating vortex dipole interaction with bottom boundary: initial condition $a)~t=0$, before first collision $(b)~t=0.25$, right after sheet with high-amplitude vorticity detaches from the boundary $(c)~t=0.36$, and before the second collision. The Reynolds number, $Re \sim 1/\nu$, equals to $Re=2500$. Element interfaces indicated for the $256 \times 256$ case with grid parameters as indicated in Table \ref{grid_parameters_vortex_xz}.}.
    \label{results_vortex_xz}
\end{figure}

For the two-dimensional case, the $x$-$z$ plane is used to define a computational domain $x \in [-1,1]$ and $ z \in [-1,1]$. Following \cite{Clercx06}, the initial condition is defined by the two-dimensional velocity field $\mathbf{u}=(u,w)$,
\begin{align}
u &= \frac{\omega_0}{2} (z - z_1) e^{(-r_1^2/r_0^2)} - \frac{\omega_0}{2} (z - z_2) e^{(-r_2^2/r_0^2)} \nonumber \\
w &= \frac{\omega_0}{2} (x - x_2) e^{(-r_2^2/r_0^2)} - \frac{\omega_0}{2} (x - x_1) e^{(-r_1^2/r_0^2)} 
\end{align}
where $\omega_0$ is the dimensionless maximum vorticity value at the center of each monopole. The location of the center of the positive and negative monopole are represented by $(x_1, z_1)=(-0.1,0)$ and $(x_2,z_2)=(0.1,0)$, with radius $r_0=0.1$. $r_1$ and $r_2$ are the radial coordinates with origin at the positive and negative monopole centers, respectively. The value of $\omega_0=299.5284$ is selected to match the parameters used in \cite{DIAMANTOPOULOS2022102065}. Vorticity contours of the initial velocity field are presented in panel $(a)$ of Fig.~\ref{results_vortex_xz}.

The total kinetic energy of the flow field,
\begin{equation}
    KE(t) = \frac{1}{2} \int_{-1}^{1}\int_{-1}^{1} \mathbf{u}^2(\mathbf{x},t) dx dz, 
\end{equation}
is normalized to be $KE(0)=2$ for the initial condition. The initial total enstrophy,
\begin{equation}
\text{Enstrophy}(t) = \frac{1}{2} \int_{-1}^{1}\int_{-1}^{1} \omega^2(\mathbf{x},t) dx dz
\end{equation}
is computed from the scalar vorticity field $\omega$ and it has an initial value of $\Omega(0)\approx 800$ for all the cases. The Reynolds number, $Re=UH/\nu$ (where $U=1$m/s is the average RMS velocity and $H=1$m is the half height of the computational domain, as in \cite{DIAMANTOPOULOS2022102065}), is increased by decreasing the kinematic viscosity $\nu$ only.

Simulations are performed for three different spatial resolutions by increasing the number of grid points in the $x$ and $z$ directions. The minimum horizontal and vertical grid spacing is reported in Table \ref{grid_parameters_vortex_xz}, where the point count is selected to match \cite{DIAMANTOPOULOS2022102065} in both directions. In $z$, a constant polynomial order of $N_p-1=7$ is used for all the choices of number of elements $N_e$. The grid is stretched vertically upward so that features within the bottom boundary layer are better resolved. The stretching strategy consists of using a constant ratio $\phi_z=h^{e+1}/ h^{e} \leq 1$ to determine the heights of successive elements. The grid points within each element are the Gauss-Lobatto-Legendre (GLL) points \cite{canuto2007spectral} which are non-equidistant. Numerical experiments have been performed for five different Reynolds numbers: $Re=625,~1250,~2500,~5000,~10000$.

\begin{table}
\begin{center}
\begin{tabular}{ c|c|c|c|c} 
  $N_x \times N_z$  & $N_e$ & $\Delta x$ & $\Delta z_{min}$ & $\phi_z$\\ 
  \hline
  $256$ $\times$ $256$ &  $32$ & $7.81 \times 10^{-3}$ & $1.55 \times 10^{-3}$ & 0.96\\ 
  \hline
  $512 \times 512$ & $64$ & $3.91 \times 10^{-3}$ & $3.31 \times 10^{-4}$ & 0.96\\ 
  \hline
  $1024 \times 1024$ & $128$ & $1.95 \times 10^{-3}$ & $2.57 \times 10^{-4}$ & 0.985\\ 
  \hline 
\end{tabular}
 \caption{Grid parameters for the simulation of the normal collision of a two-dimensional vortex dipole with a no-slip wall. The number of modes, $N_p$, is set to be $N_p=8$ for all the cases (polynomial order $N_p-1=7$), with varying vertical grid stretching factor $\phi_z$. The number of grid points in the $x$ direction is $N_x$ and the number of grid points in the $z$ direction is $N_z$ ($N_z = N_e \times N_p$). Grid stretching constant in the $z$ direction is denoted by $\phi_z$, and the minimum grid spacing in each direction is denoted by $\Delta x$ and $\Delta z_{min}$.}
 \label{grid_parameters_vortex_xz}
\end{center}
\end{table}
Contour plots of the vorticity field, before and after the first collision, are presented in Fig.~\ref{results_vortex_xz}. As expected, a boundary layer is formed at the bottom boundary as the dipole approaches it. After the first collision occurs, a high-amplitude vorticity region detaches from the boundary and contributes to forming an asymmetric dipole (see Fig.~\ref{results_vortex_xz}$(c)$). The asymmetric dipole moves away from the boundary following a curved trajectory that approaches the boundary again before a second collision takes place (see Fig.~\ref{results_vortex_xz}$(d)$). 

\begin{figure}
\centering
\subfloat[$Re = 1250$\label{fig:1a}]{\includegraphics[scale=0.24, trim={7cm 2.8cm 7cm 3cm},clip]{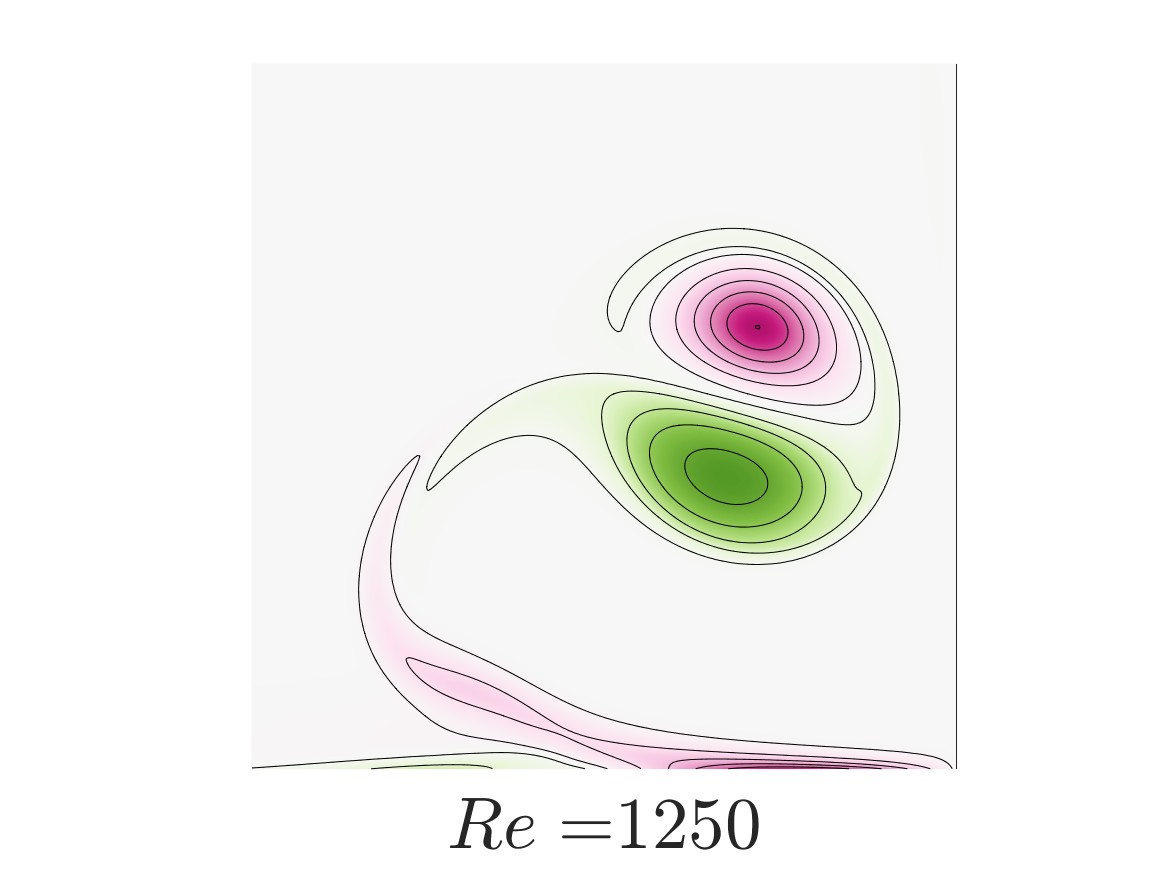}}
\subfloat[$Re = 2500$\label{fig:1b}] {\includegraphics[scale=0.24, trim={7cm 2.8cm 7cm 3cm},clip]{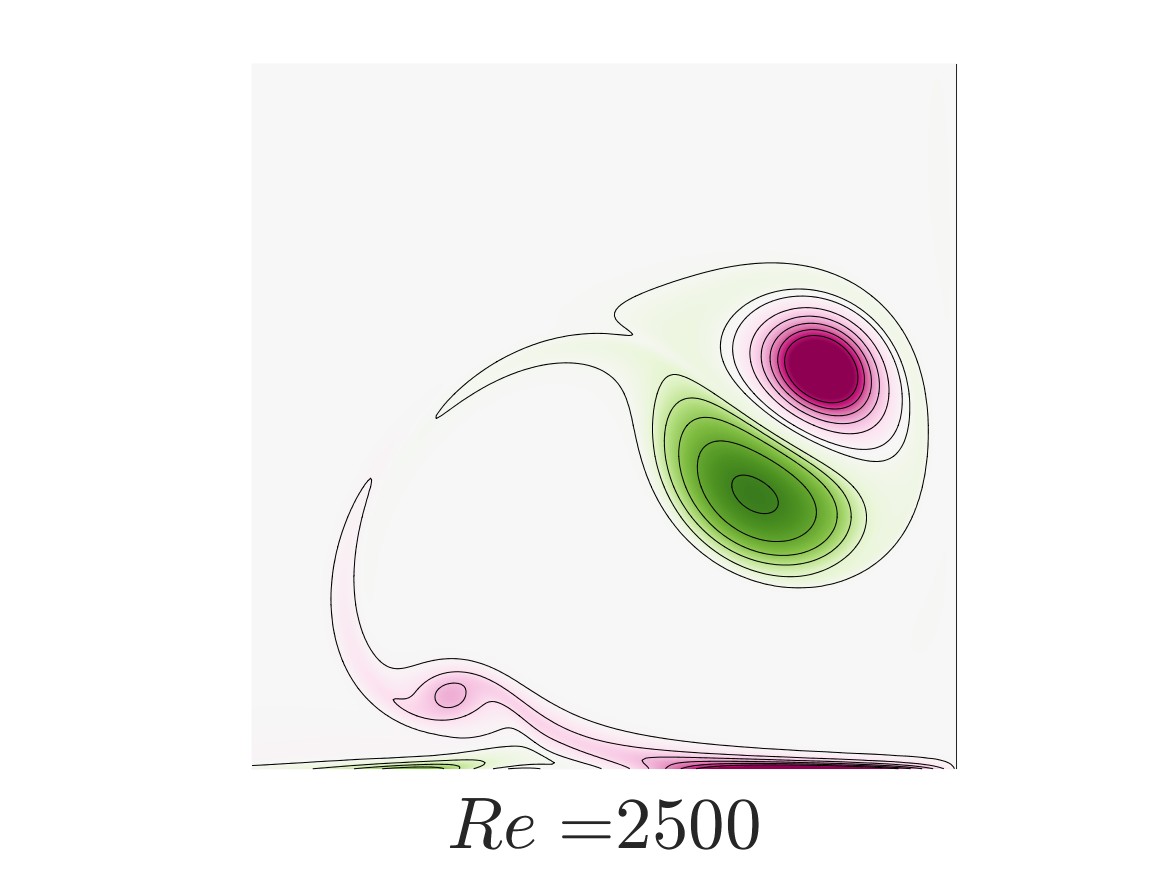}
} \hfill
\subfloat[$Re = 5000$\label{fig:1c}]{\includegraphics[scale=0.24, trim={7cm 2.8cm 7cm 3cm},clip]{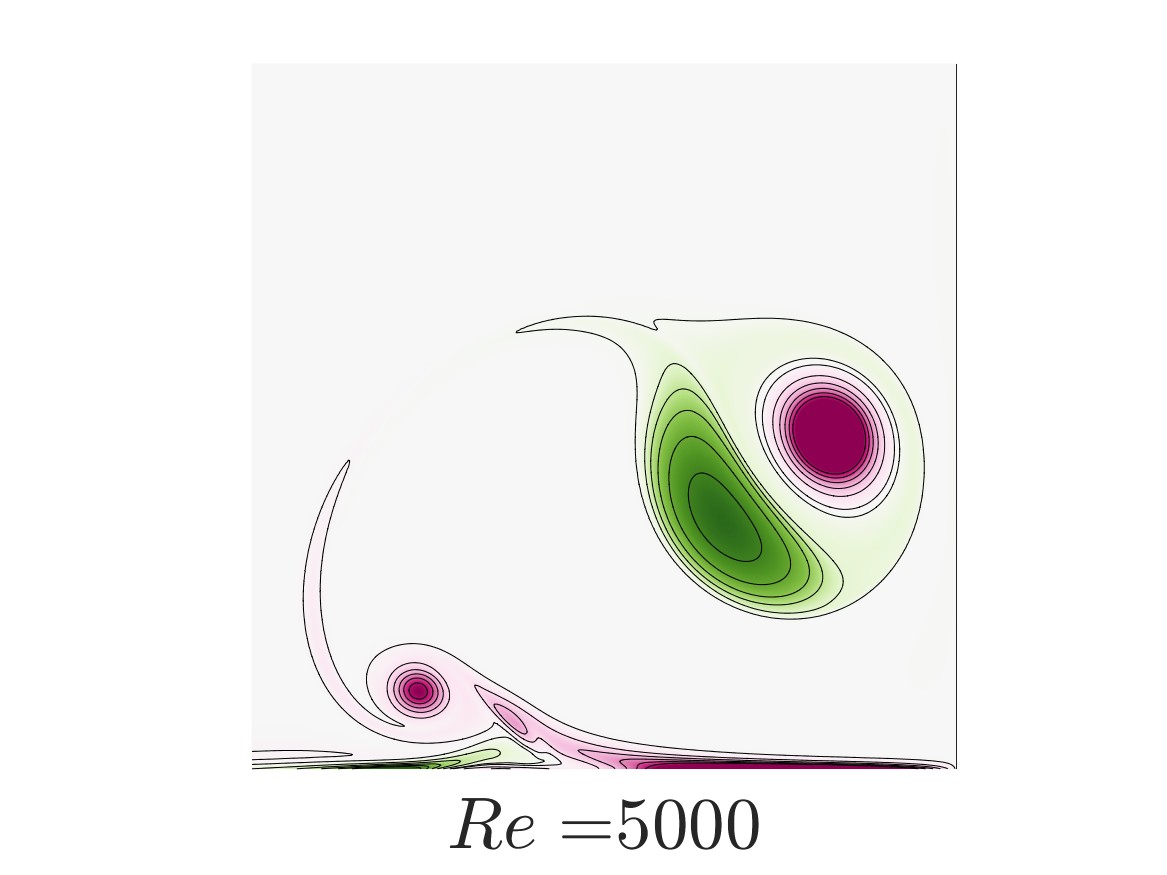}} 
\subfloat[$Re = 10000$\label{fig:1d}]{\includegraphics[scale=0.24, trim={7cm 2.8cm 7cm 3cm},clip]{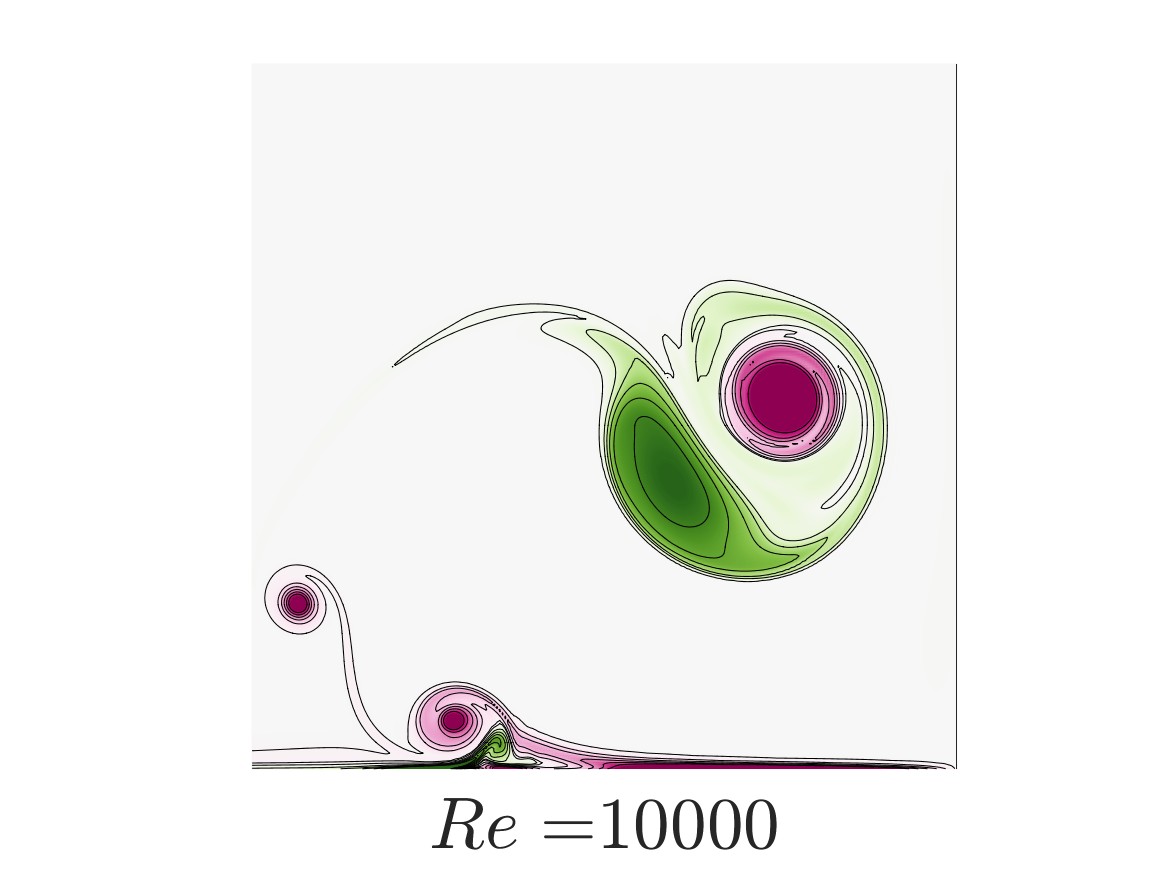}
}

\caption{Vorticity contour plots of the dipole-wall collision at $t=0.51$ for $Re=1250$ (a), $2500$ (b), $2500$ (b), $5000$ (c), and $10000$ (d). Only a fraction of the computational domain is presented, with $-0.5 \leq x \leq 1.0$ and $-1.0 \leq z \leq 0.0$. Spatial resolution corresponds to $1024 \times 1024$ case from Table \ref{grid_parameters_vortex_xz}. Contour levels of non-dimensional vorticity in the range $\{-260, 260 \}$ values drawn every $50$, where pink colors indicate negative values, and green colors positive ones. As $Re$ grows, the thickness of the vorticity filament at the bottom boundary decreases. At $x=-0.2$, the estimated thickness of the filament is: (a)~$25.6\times 10^{-3}$, (b)~$19.5\times 10^{-3}$, (c)~$14.0\times 10^{-3}$, and (d)~$10.10\times 10^{-3}$. In the latter, the spatial resolution guarantees at least $14$ grid points, in the vertical direction, to represent this filament (specifically at $x=-0.2$).} \label{vortex_zoom_Re}
\end{figure}

As the Reynolds number increases, the magnitude of the vorticity generated at the boundary grows, and additional small vortices and vorticity filaments are created. Fig.~\ref{vortex_zoom_Re} shows the vorticity contours of the positive signed primary vortex before the second collision takes place for different $Re$ values. The presence of smaller-scale flow structures with increasing $Re$ is evident. However, the trajectory of the positive signed primary vortex is barely impacted by the Reynolds number, in agreement with the observations reported by \cite{Clercx06}. 

\begin{figure}
     \centering
      \includegraphics[scale=0.3]{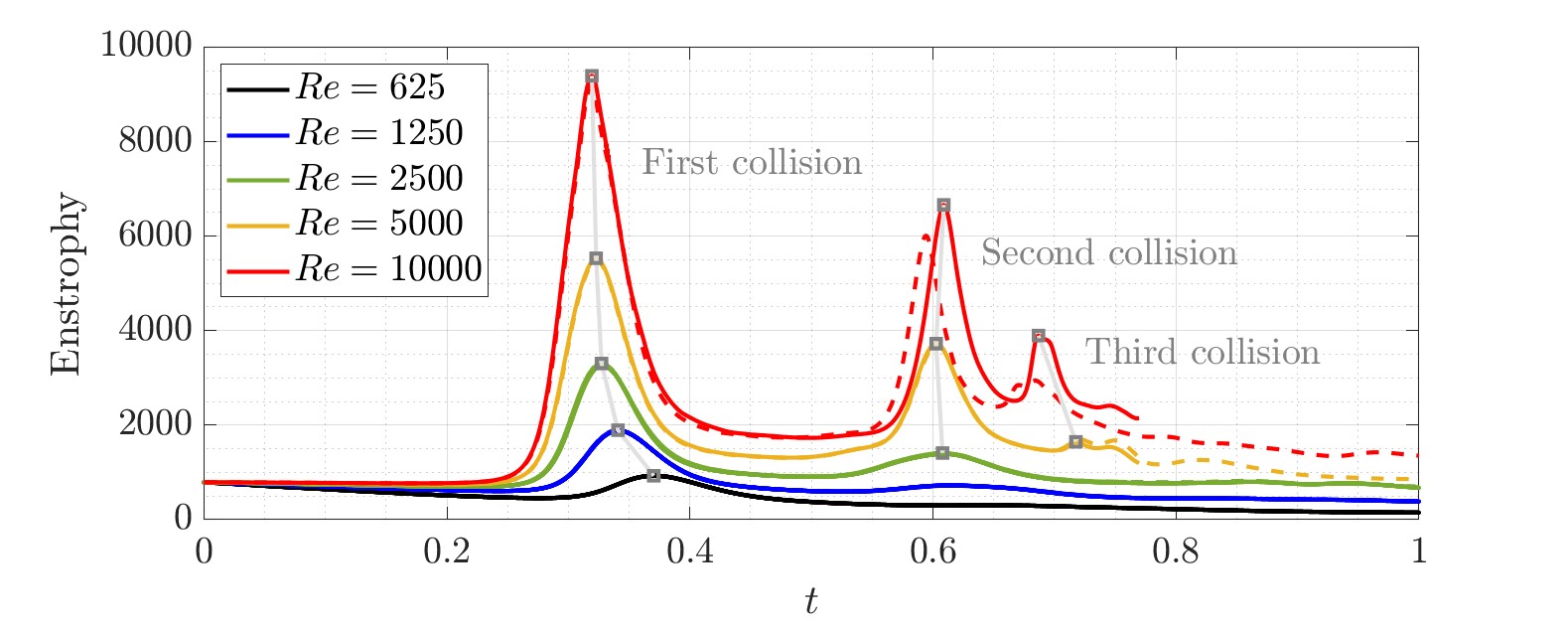}
     \caption{Total enstrophy vs time for the vortex dipole test case different $Re$ and spatial resolutions: $256 \times 256$ (dotted line), $512 \times 512$ (dashed line) and $1024 \times 1024$ (solid line). The collisions of the primary vortex with the no-slip wall generate peaks in the evolution of the Enstrophy.}
     \label{fig:enstrophyvsTime}
\end{figure}

The simulation results show that every collision of the primary vortex with the no-slip wall generates an increase in the vorticity production at the bottom boundary, and therefore, a peak in the total enstrophy value (as reported by \cite{Clercx06}). These peaks are visible in the time-series plot of Fig.~\ref{fig:enstrophyvsTime}, where the evolution of the total enstrophy is presented for several $Re$ and the spatial resolutions from Table \ref{grid_parameters_vortex_xz}. The figure indicates the times of the first and second collisions, which are more intense for the highest $Re$ cases. A lower-intensity third collision can only be distinguished for the two highest $Re$ values. 

\begin{figure}
     \centering
          \includegraphics[scale=0.38]{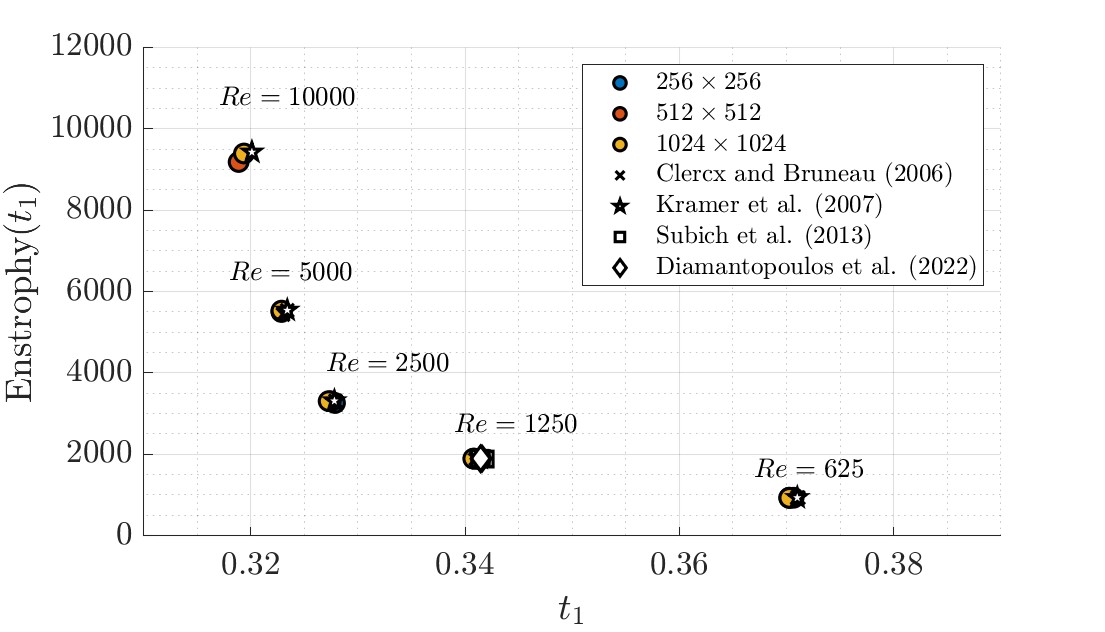}
 \caption{Maximum total enstrophy during the first collision of a vortex dipole with a no-slip wall, and its time of occurrence ($t_1$) for different grid resolutions and $Re$. Values reported by other studies are shown for comparison (\cite{Clercx06,Kramer07,Subich_methods_2013,DIAMANTOPOULOS2022102065})}
 \label{maxvalsEnstroPlot}
\end{figure}

\begin{figure}
     \centering
          \includegraphics[scale=0.32]{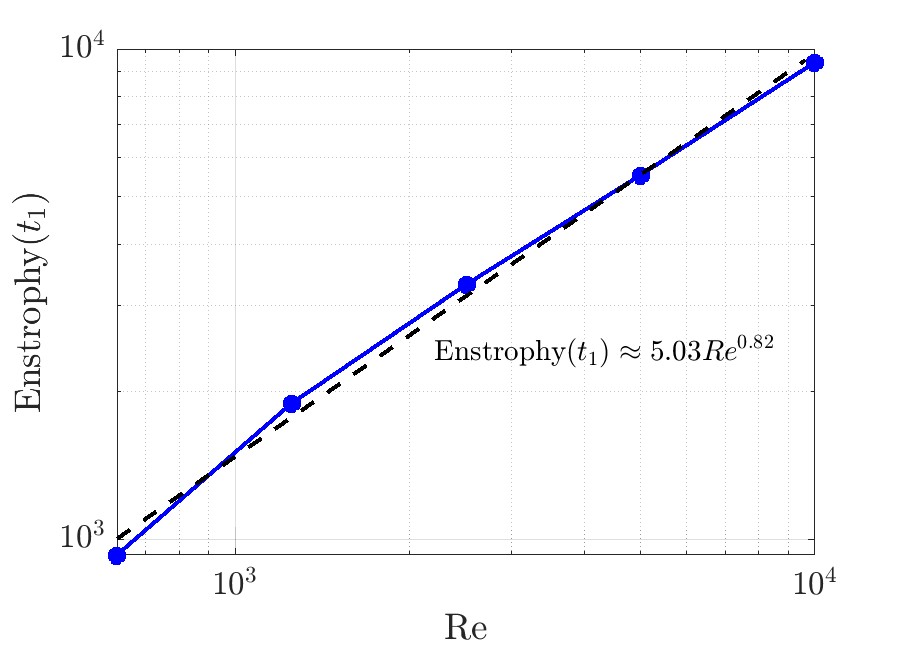}
 \caption{Maximum total enstrophy during the first collision ($t_1$) of a vortex dipole for increasing $Re$. Blue markers correspond to the $1024 \times 1024$ spatial resolution results from Fig.~\ref{maxvalsEnstroPlot}. Black dashed line is a linear fit obtained through least-squares regression. Values scale as Enstrophy$(t_1)\sim Re^{0.8}$, in agreement with results reported by \cite{Clercx02}.}
 \label{FitEnstroRe}
\end{figure}

To offer a broader summary of the results of the flow solver, Fig.~\ref{maxvalsEnstroPlot} compares the first enstrophy peak and its time of occurrence against values reported by previous studies (\cite{Clercx06, Kramer07, Subich_methods_2013, DIAMANTOPOULOS2022102065}). The latter studies emply numerical discretizations based on pseudospectral Chebyshev methods (\cite{Clercx06}), Fourier/Chebyshev expansions (\cite{Kramer07, Subich_methods_2013}), and spectral element/Fourier-Galerkin schemes (\cite{DIAMANTOPOULOS2022102065}). The results of the flow solver show the expected variation of enstrophy with respect to $Re$. For instance, Fig.~\ref{FitEnstroRe} confirms that the maximum enstrophy at the time of the first collision scales as Enstrophy$(t_1)\sim Re^{0.8}$, as reported by \cite{Clercx02}. The enstrophy values and collision times associated with the first collision are detailed in Table \ref{maxvalsEnstro}. 

The accuracy of the enstrophy peak and time of occurrence increases with finer spatial resolution. For example, a finer grid produces differences in the results in the $Re=10000$ case (see Fig.~\ref{fig:enstrophyvsTime}). Such results approach the values reported by \cite{Kramer07} as the grid point count increases from $512 \times 512$ to $1024 \times 1024$ points. As a reference, \cite{Kramer07} used an even finer grid, $1536 \times 1024$ points, to simulate the latter $Re$ case. Given that the flow solver reproduces the expected flow dynamics for this Navier-Stokes benchmark, as well as the differences in the enstrophy as $Re$ increases, the differences in values concerning other studies are mainly attributed to time step and actual grid resolution.

\begin{table}[ht]
\begin{center}
\begin{tabular}{ l|l|l|l} 
 $Re$ & $N_x \times N_z$ & $t_1$ & Enstrophy$(t_1)$ \\ 
 \hline
 625 &  256 $\times$ 256 & 0.37066 & 929.84 \\
     &  512 $\times$ 512 & 0.37026 & 929.52 \\
     &  1024 $\times$ 1024 & 0.37026 & 928.99 \\
     &  Clercx and Bruneau (2006)\cite{Clercx06}  & 0.3711 & 933.60 \\   
 \hline
 1250 & 256 $\times$ 256 & 0.34126 & 1881.76\\
     &  512 $\times$ 512 & 0.34086 & 1892.62 \\ 
     &  1024 $\times$ 1024 & 0.34076 & 1892.53 \\
     &  Clercx and Bruneau (2006) \cite{Clercx06}  & 0.3414 & 1899 \\
     &  Subich et al. (2013)\cite{Subich_methods_2013}  & 0.34137 & 1899.21 \\
     &  Diamantoupulos (2022)\cite{DIAMANTOPOULOS2022102065}  & 0.34145 & 1897.24 \\ 
 \hline
 2500 &  256 $\times$ 256 & 0.32786 & 3259.79 \\ 
      &  512 $\times$ 512 & 0.32746 & 3299.69 \\ 
      &  1024 $\times$ 1024 & 0.32726 & 3305.64 \\
      &  Clercx and Bruneau (2006)\cite{Clercx06}  & 0.3279 & 3313 \\ 
 \hline
 5000 &  512 $\times$ 512 & 0.32286 & 5497.45 \\
      &  1024 $\times$ 1024 & 0.32286 & 5530.69 \\ 
      &  Clercx and Bruneau (2006)\cite{Clercx06}  & 0.3234 & 5536 \\
      &  Kramer et al. (2007) \cite{Kramer07}      & 0.3234 & 5540 \\ 
 \hline
 10000 &  512 $\times$ 512 & 0.31886 & 9182.59 \\ 
       &  1024 $\times$ 1024 & 0.31936 & 9392.22 \\ 
       &  Kramer et al. (2007) \cite{Kramer07}& 0.3201 & 9426
\end{tabular}
 \caption{Maximum total enstrophy and its time of occurrence for different grid resolutions and $Re$. Values reported by other studies are shown for comparison (\cite{Clercx06, Subich_methods_2013,DIAMANTOPOULOS2022102065, Kramer07})}
 \label{maxvalsEnstro}
\end{center}
\end{table}

Lastly, note that the findings resulting from this series of collisions of a dipole with a no-slip wall are useful to validate the flow solver, but may not have a physical meaning due to its two-dimensionality. In a three-dimensional setting, the vorticity production at the no-slip boundary is expected to be more intense, with flow structures and vortex trajectories qualitatively different due to mutual interactions in the added transverse direction.

\subsubsection{Three-dimensional effects}
\begin{figure}
     \centering
         \centering
          \includegraphics[scale=0.33, trim={3cm 5cm 0cm 4cm},clip]{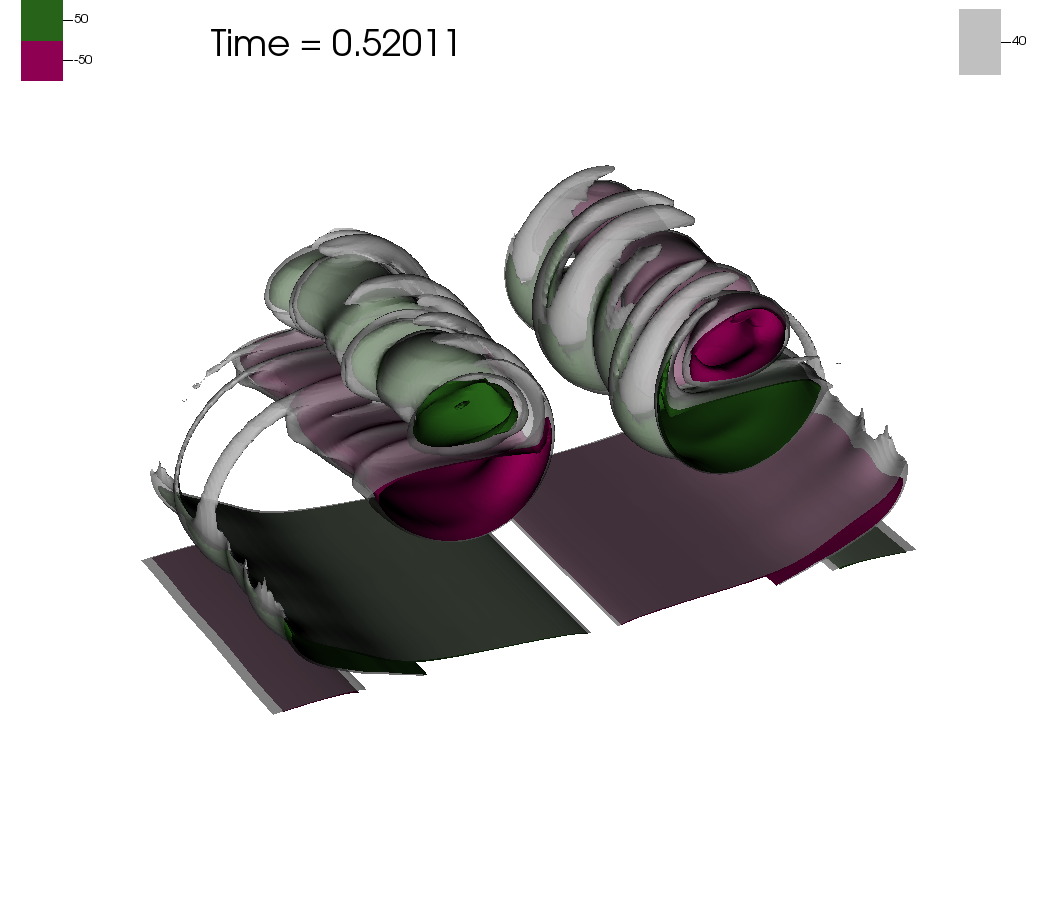}
         \includegraphics[scale=0.33, trim={3cm 5cm 0cm 10cm},clip]{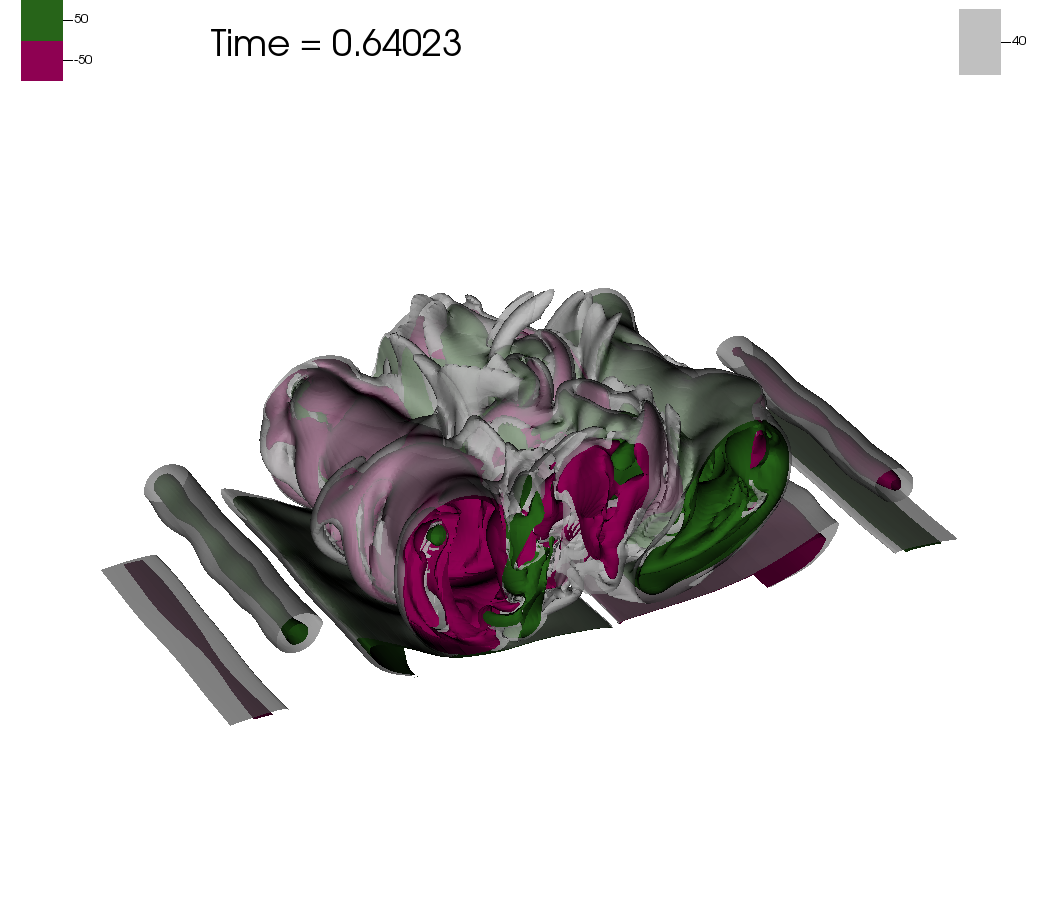}
         \caption{Isosurfaces of the transverse vorticity $\omega_y=-50$ (pink) and $\omega_y={50}$ (green), and magnitude of the vorticity $|\omega|=40$ in gray, before the second collision with the no-slip wall at $t=0.51$ (top), and right after the collision takes place, at t=0.61 (bottom). $\omega=10^{-3}$ test case.}
         \label{fig:3DviewDipole}
\end{figure}

Following the same approach as \cite{Subich_methods_2013}, the simulation of the vortex dipole collision can be extended to a three-dimensional domain when Gaussian noise is added to the initial velocity field. The flow structure generated at the wall undergoes a three-dimensional instability and can transition to turbulence. To assess the impact of the amplitude of the noise, as proposed by \cite{DIAMANTOPOULOS2022102065}, two values of standard deviation ($\sigma$) of the white noise are evaluated. The Gaussian white noise standard deviation values correspond to $\sigma:10^{-2},~10^{-3}$. To illustrate the phenomenon, Fig.~\ref{fig:3DviewDipole} shows an isosurface of the magnitude of the vorticity $|\boldsymbol{\omega}|=\Big( \omega_x^2 + \omega_y^2 + \omega_z^2 \Big)^{1/2}$, as well as two isosurfaces of the transverse vorticity, $\omega_y$, for the $\sigma=10^{-3}$ case. The vorticity structures in the transverse direction indicate the three-dimensionalization of the flow. Note that the selected spatial resolution, for the $x$ and $z$ directions, corresponds to $N_x \times N_z=512 \times 512$ (see Table~\ref{grid_parameters_vortex_xz}). In $y$, the computational domain defined by $y \in [0,0.4]$ is discretized using $96$ grid points and a uniform grid spacing $\Delta y=4.16\times 10^{-3}$. 

Differences in the evolution of kinetic energy and enstrophy are expected for the different $\sigma$ values. For the stronger perturbations ($\sigma=10^{-2}$), a faster (\textit{bypass}) transition to turbulence is anticipated, as well as the formation of richer three-dimensional fine-scale structures in the flow. In contrast, for the weaker perturbations ($\sigma=10^{-3}$), the transition to turbulence and formation of small three-dimensional structures is expected to develop at a slower rate, with a lower production of vorticity at the bottom boundary, and less kinetic energy dissipation due to the lack of three-dimensional flow structures. 

\begin{figure}
     \centering
         \centering
          \includegraphics[scale=0.22]{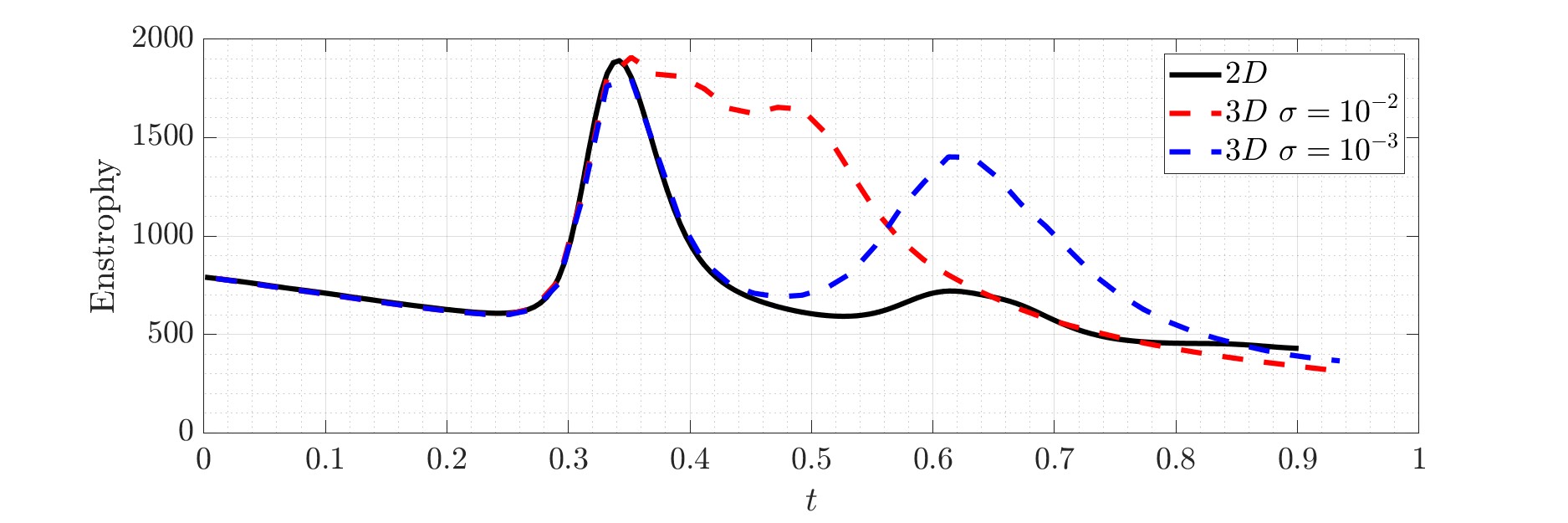}
           \includegraphics[scale=0.22]{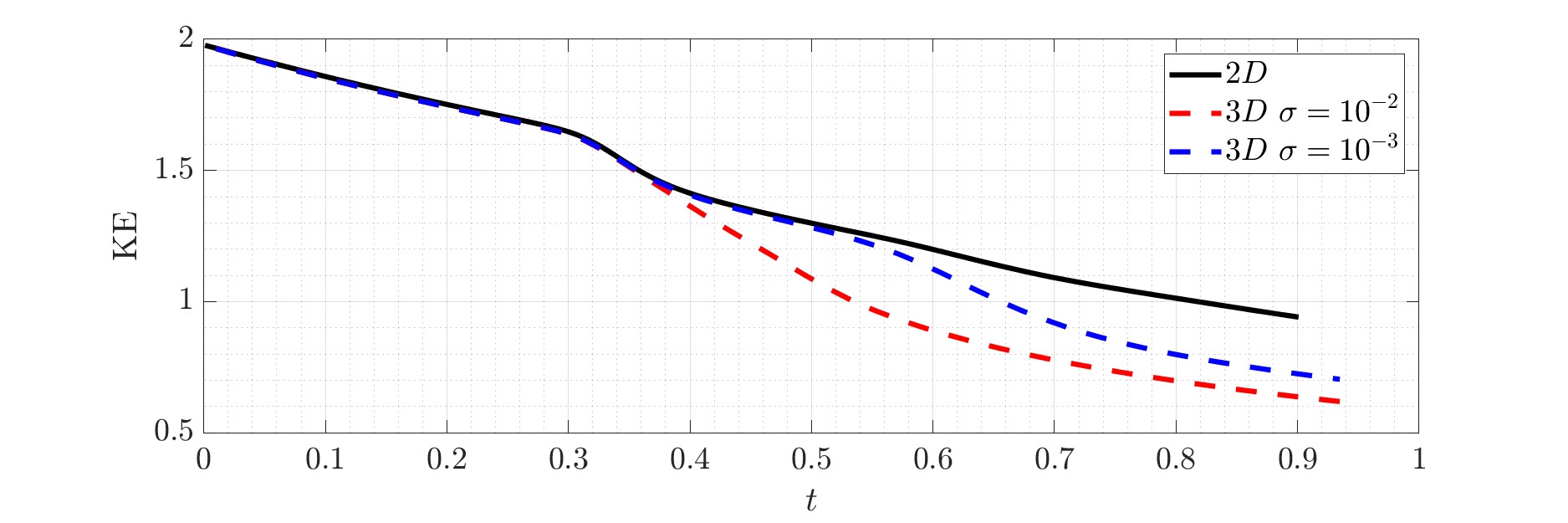}

         \caption{Total enstrophy vs time (top) and total kinetic energy (bottom) for the vortex dipole test case $Re=1250$. Solid line corresponds to the two-dimensional case with spatial resolution $1024 \times 1024$. Dotted lines correspond to the three dimensional case, where white Gaussian noise of a $10^{-2}$ standard deviation (red) and $10^{-3}$ have  been added to the initial condition. Enstrophy and kinetic energy have been averaged in the $y$ direction.}
         \label{fig:enstrophyvsTime_3D}
\end{figure}

The simulation results reflect the expected dynamics. The $y$-average enstrophy and kinetic energy follow the same behaviour reported by \cite{DIAMANTOPOULOS2022102065}, where the three-dimensional cases resemble the behaviour of the two-dimensional case until the primary vortex collides against the no-slip wall for the first time (see Fig.~\ref{fig:enstrophyvsTime_3D}). At this point, the enstrophy for the different values of $\sigma$ shows a different evolution. The weaker perturbation case ($\sigma = 10^{-3}$) continues to follow the two-dimensional benchmark until the transverse flow structure grows, before the second collision, and generates a stronger peak of enstrophy of the same order of magnitude as the first one. For the stronger perturbations ($\sigma = 10^{-2}$), the three-dimensionalization starts right after the first collision takes place, earlier than in the $\sigma = 10^{-3}$ case, and a decay in kinetic energy is observed earlier than with $\sigma = 10^{-3}$ (see bottom of Fig.~\ref{fig:enstrophyvsTime_3D}). 

\begin{figure}
     \centering
         \centering
          \includegraphics[scale=0.30]{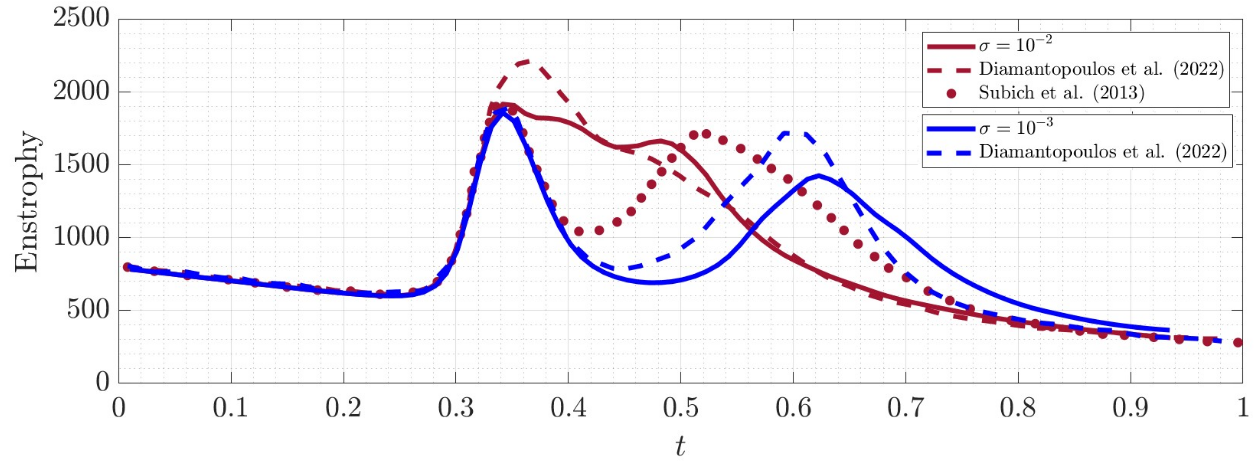}
         \caption{Evolution of the $y$-averaged total enstrophy for the three-dimensional vortex dipole with $Re=1250$. Gaussian noise with two standard deviation ($\sigma$) values is added to the initial velocity field: $\sigma=10^{-2}$ (red color lines) and $\sigma=10^{-3}$ (blue color lines). Simulation results (solid lines) are compared against other numerical studies: \cite{DIAMANTOPOULOS2022102065} (dashed lines) and \cite{Subich_methods_2013} (dotted line).}
         \label{fig:enstrophy3Dcomp}
\end{figure}

Specifically for the three-dimensional case, Fig.~\ref{fig:enstrophy3Dcomp} shows a comparison of the $y$-averaged enstrophy against the results reported by \cite{Subich_methods_2013} and \cite{DIAMANTOPOULOS2022102065}. The evolution of the enstrophy is similar to the one indicated by \cite{Subich_methods_2013}, although the peak values are less intense than reported by \cite{DIAMANTOPOULOS2022102065}. These variations are attributed to differences in the structure of the noise and the spatial resolution used in the $x$ direction, which is finer, particularly in \cite{DIAMANTOPOULOS2022102065}, where $\Delta x_{min} = 3.249 \times 10^{-4}$.

\subsection{Tankscale ISW in uniform depth water}
The propagation of a tankscale internal solitary wave (ISW) in uniform depth water, with a continuous two-layer stratification, assesses the numerical dissipation and dispersion of the solver and its ability to simulate strongly nonlinear and non-hydrostatic effects. The numerical scheme has to be able to correctly reproduce the sensitive balance between nonlinear steepening and physical dispersion. Moreover, the wave amplitude should not artificially attenuated by numerical diffusion \cite{StasnaIWs22, DIAMANTOPOULOS2022102065} and no spurious trailing waves should emerge due to numerical dispersion. 

The initial wave is an exact solution of the Dubreil-Jacotin-Long (DJL)  equation. The latter is a nonlinear eigenvalue problem which produces a wavefield that is an exact solution of the full Euler equations under the Boussinesq approximation. A reliable incompressible Euler solver, with a DJL generated ISW as initial condition, should reproduce a wave that honors the above balance between nonlinearities and dispersion, and maintains a constant waveform, propagation speed and wave-integrated kinetic energy. As such, this benchmark is purely inviscid and non-diffusive mode.  

\begin{figure}
    \centering
    \includegraphics[scale=0.30]{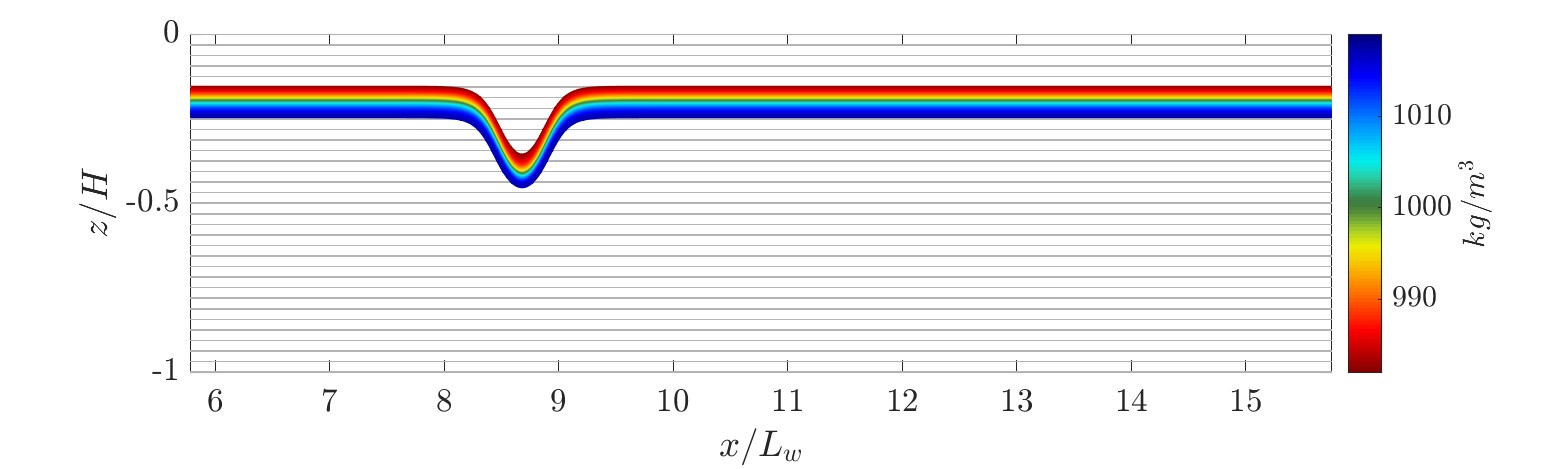} 
    \caption{Isopycnal contours of the total density for used as the initial condition for the ISW inviscid case. Vertical element interfaces indicated according to the grid parameters as listed in Table \ref{grid_parameters_ISW}.}
     \label{fig:density_ISW}
\end{figure}

As in \cite{DIAMANTOPOULOS2022102065}, the two-layer stratification profile is described by,
\begin{equation}
\rho = \rho_0 - \frac{\Delta \rho}{2}\tanh\big( \frac{z+h}{\delta} \big)
\end{equation}
where $\Delta \rho = 40~\mathrm{kg/m^3}$ is the difference in density between the two layers, $h=3~\mathrm{cm}$ is the upper layer depth and $\delta = 0.5~\mathrm{cm}$ is the interface thickness. The reference density, $\rho_0$, is equal to $1000~\mathrm{kg/m^3}$ and the initial available potential energy of the ISW is $0.05~\mathrm{J/m}$. Per the above discussion, the wave is expected to propagate at a speed of $c=0.1145~\mathrm{m/s}$, preserving its initial width, $L_w$, which is set to be $L_w \approx 0.69~\mathrm{m}$. The streamwise two-dimensional domain has a total length of $L = 10 \times L_w = 6.9~\mathrm{m}$ and a depth of $H=15~\mathrm{cm}$. Isopycnal contours of the ISW used as initial condition are generated using the method of \cite{Dunphy2011} and presented in Fig.~\ref{fig:density_ISW}.  

Table \ref{grid_parameters_ISW} shows the number of grid points in each direction and the minimum horizontal and vertical grid spacing used in the simulations. The number of points was chosen to match the setup used in \cite{DIAMANTOPOULOS2022102065}. All elements have the same length and use a seventh-order polynomial ($N_p-1=7$) within each element. We performed different numerical experiments varying the intensity of the spectral filtering, which is designed to preserve numerical stability (see Appendix A) and controls, for the most part, the numerical diffusion in the flow solver. A series of exponential filter order values: $32,~20,~16,~14,~12$, were evaluated to assess their effects on the magnitude of the numerical dissipation. In the order listed, they vary from a very weak filter function, where numerical dissipation is expected to be minimal and restricted to the smallest length scales, to a strong one, where numerical dissipation is anticipated to have an effect on a broader range of length scales in the flow.   

\begin{table}
\begin{center}
\begin{tabular}{ c|c|c|c|c } 
  $N_x \times N_z$  & $N_e$ & $\Delta x$ & $\Delta z_{min}$ & $\phi_z$\\ 
  \hline
  $512$ $\times$ $256$ &  $32$ & $1.35 \times 10^{-2}$ & $8.14 \times 10^{-5}$ & $1.0$
\end{tabular}
 \caption{Grid parameters for the inviscid tank-scale ISW. Polynomial order $N_p-1=7$ is constant with uniform size elements. Number of grid points in the $x$ direction is $N_x$, and number of grid points in the $z$ direction is $N_z$. Vertical grid stretching factor $\phi_z$. Minimum grid spacing in each direction is denoted by $\Delta x$ and $\Delta z_{min}$}.
 \label{grid_parameters_ISW}
\end{center}
\end{table}

\begin{figure}
    \centering
    \includegraphics[scale=0.25]{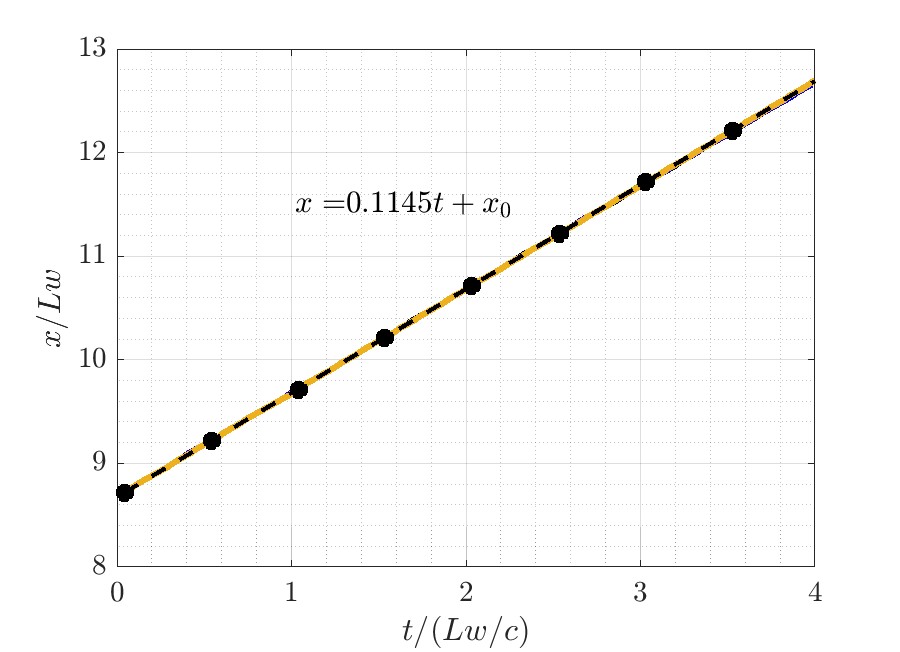} 
    \includegraphics[scale=0.25]{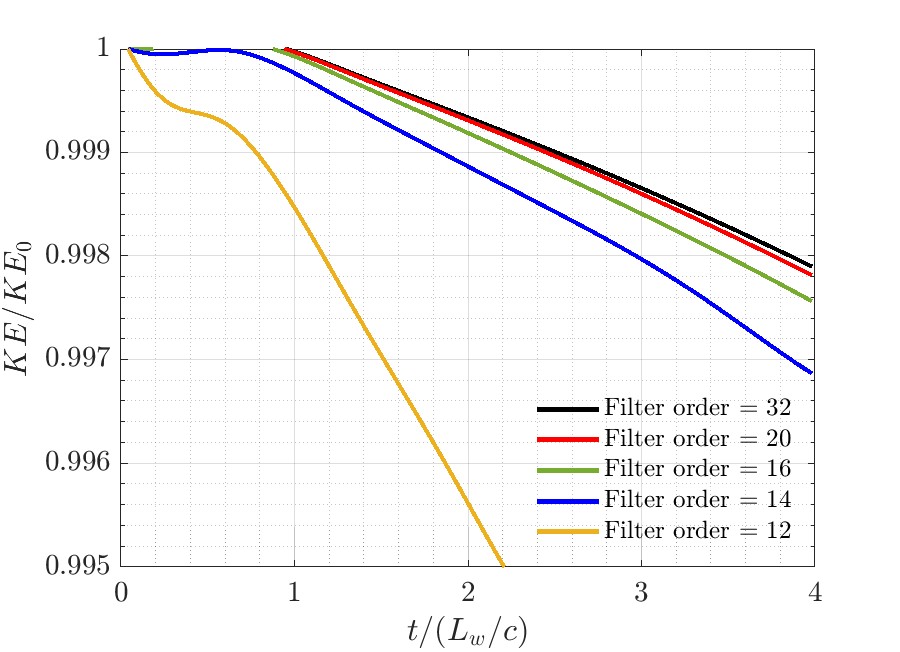} 
    \caption{Position of the wave (left) and normalized, by its initial values, total kinetic energy per unit volume vs time (right) for the ISW benchmark problem. The ISW propagates for four wavelengths $L_w$. On the left, the position of the wave vs time (yellow color line) corresponds to a numerical experiment with filter order equals to $12$. A linear fit of the results (dashed-dotted black color line) allows us to estimate the propagation speed of the wave to be $c_{\text{approx}}=0.1144774~\mathrm{m/s}$. On the right, kinetic energy results are shown for numerical experiments with varying filter order value. Larger filter order values indicate a weaker filter.}
    \label{fig:KE_ISW}
\end{figure}

The simulated ISW indeed maintains its shape as it travels at a constant speed. Using a linear fit of the location of the wave over time, we estimate its propagation speed to be $c_{\text{approx}}=0.1144774~\mathrm{m/s}$ (see left panel in Fig.~\ref{fig:KE_ISW}). In the numerical experiments we performed, the propagation speed differed from the theoretical prediction by less than $0.1\%$. The latter speed did not change when the exponential filter order value was varied, indicating that the filtering procedure barely influenced the mean velocity scales in the flow. Furthermore, at every instant of time, the kinetic energy, KE, is computed in the two-dimensional domain $\Omega$ as,
\begin{equation}
KE = \frac{\rho_0}{2} \int_{\Omega} \mathbf{u}^2 (\mathbf{x},t) d \Omega.
\end{equation}

In the absence of viscous dissipation, $KE$ is expected to remain constant. Any deviations from its initial value are attributed to numerical dissipation of the flow solver. The evolution of the normalized---by its initial values---KE, as a function of time and filter order, is reported in Fig.~\ref{fig:KE_ISW}. As expected, a stronger numerical filter increases the rate at which the KE decreases. After two characteristic wave-propagation time scales, the KE can decrease from $0.05\%$ to $0.5\%$---depending on the filter order. The slight decrease in kinetic energy observed in the simulations helps us confirm that the flow solver introduces minimal numerical dissipation, which is found to be comparable in magnitude to what is reported by \cite{DIAMANTOPOULOS2022102065}.


\subsection{Stratified turbulent wakes} \label{sec:simsetup}
A stratified turbulent wake with non-zero net momentum is used to evaluate the capabilities of the flow solver to reproduce highly turbulent flow developing in a stably stratified fluid. Experimental work and wake theory for a towed sphere wake, as well as other numerical studies, allow us to evaluate the correctness of the solver by comparing flow features and power decay laws, particularly for the mid stages of the wake evolution. An overview of the validation approach and simulation setup is presented below.
\subsubsection{Simulation setup}

\begin{figure}[ht]
     \centering
         \centering
          \includegraphics[scale=0.36]{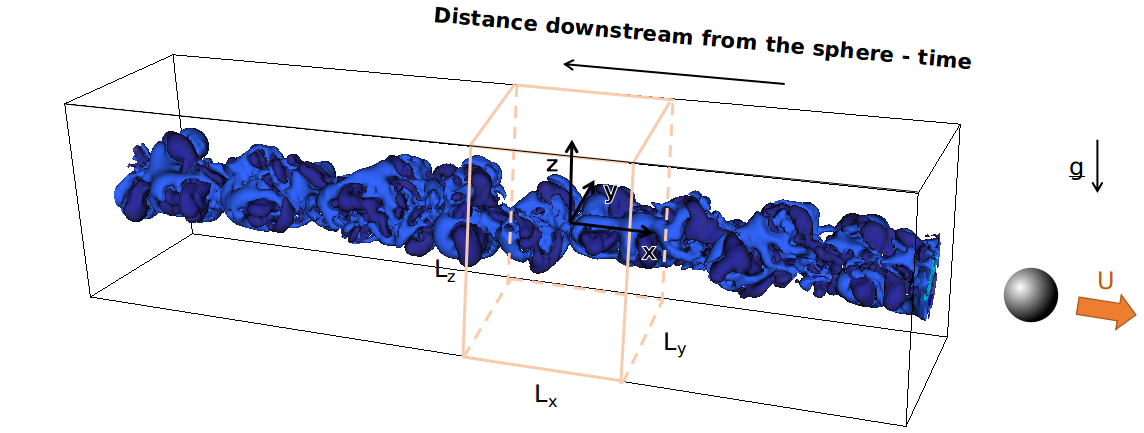}
         \caption{Schematic of simulation setup for sphere wake simulations, as proposed by \cite{Dommermuth2002}. Only a portion of the wake is simulated in the computational domain (orange box) at a time. Sphere of diameter $D$ moves in the $x$ direction at constant speed $U$. Length of the computational domain is much shorter than the total wake length.}
         \label{fig:wakeSetUp}
\end{figure}

The setup is similar to the one used by \cite{diamessis_spedding_domaradzki_2011} and \cite{ZhouDiamessis2019}, and it consists of a three-dimensional box of dimensions $L_x \times L_y \times L_z = 8D \times 2D \times 1.8D$, where $D$ is the sphere diameter. The sphere has moved through the box from left to right at a constant speed $U$, and it is not included in the simulation (see diagram in Fig.~\ref{fig:wakeSetUp}). Given that the total wake length is considered too long to simulate in a single computational domain, the flow is simulated in a box of size $L_x \times L_y \times L_z$, shorter than the total wake length. The simulation approach follows \cite{Dommermuth2002,ORSZAGPao1975}, where the wake is considered statistically homogeneous in $x$, and \textit{time evolution of the flow is interpreted as the downstream variation of the wake} \cite{ORSZAGPao1975}. Under this assumption, selecting a computational domain length $L_x$ much smaller than the total wake length allows us to use periodic boundary conditions in the $x$ direction. We also assume periodic boundary conditions in the $y$ direction, where the width $L_y$ is chosen to be large enough to prevent interactions of the wake with its own periodic image. The reader is referred to \cite{ORSZAGPao1975,zhou_diamessis_2016} for additional details on the wake simulation setup. 

Note that $L_x$ is selected to double the size used in \cite{ZhouDiamessis2019},  as a way of attending to one of their recommendations regarding the need to improve late-time statistics of the simulated turbulent flow. A longer domain is particularly critical for the $Re \sim O(10^6)$ case, where large scale flow structures with layered turbulence are expected at late times (during the SSR), requiring a domain long enough to capture their length scale while satisfying the assumption of statistical homogeneity in $x$. 

 In these body-exclusive simulations, the mean and turbulent velocity fields at a short distance downstream from the sphere are approximated to build the initial condition; these approximations are based on the assumption of an axisymmetric wake \cite{ORSZAGPao1975, Meunier2006, diamessis_spedding_domaradzki_2011}. Details on the initialization process, and linking the particular initial mean/fluctuating velocity profiles to a sphere with diameter $D$ and speed $U$, are described in \cite{Nidiathesis}. The body-based $Re$ and $Fr$ are computed as follows,
\begin{equation}
    Re = \frac{UD}{\nu},
\end{equation}
\begin{equation}
    Fr = \frac{U}{ND}.
\end{equation}
The magnitude of the Reynolds number is varied by decreasing the kinematic viscosity coefficient $\nu$, and the Froude number $Fr$ by modifying the value of the stratification frequency $N$ that describes the background density profile $\bar{\rho}$ in the vertical direction, $\bar{\rho}(z) = -N^2 (\rho_0/g) z$.

Numerical experiments have been performed for three different Reynolds numbers with and without the influence of stratification. Simulations will be presented for three values of body-based $Re$: $Re=5 \times 10^3,~1\times 10^5,~4\times 10^5$. We use a linear stratification profile with a $Fr$ equal to $4$. The number of grid points in each direction and minimum grid spacing are reported in Table \ref{grid_parameters_wakes}. Again, $Re$ is increased by decreasing the value of the kinematic viscosity $\nu$. Note that the temporal evolution of the velocity fields is presented in terms of non-dimensional $Nt$ units, where $N$ is the Brunt–Väisälä or stratification frequency, and $t$ is time.

For the present, we have obtained results of the simulation at a higher body-based $Re$, i.e. $Re=1.6 \times 10^6$, until the early stages of the stratified simulation ($Nt \approx 12$) only, given the large demand of computational resources involved. The reader is referred to \cite{Nidiathesis} for an overview of the latter results. Stratified simulations for lower $Re$ values have been obtained for longer integration times ($Nt \approx 50$). A summary of the results is offered in the next section. 

\begin{table}[ht]
\begin{center}
\begin{tabular}{ c|c|c|c} 
    & $Re=5.0\times10^3$ & $Re=1.0\times10^5$ & $Re=4.0\times10^5$ \\
   \hline \hline
  $N_x \times N_y \times N_z$  & $512\times128\times234$ & $1024\times256\times767$ & $2048\times512\times1417$ \\ 
  \hline
  $N_e$ & $18$ & $60$ & $109$\\
  \hline
  $\Delta x/D$ & $1.04 \times 10^{-1}$ & $5.21 \times 10^{-2}$ & $2.60 \times 10^{-2}$\\
  \hline
  $\Delta y/D$   & $1.04 \times 10^{-1}$ & $5.21 \times 10^{-2}$ & $2.60 \times 10^{-2}$ \\
  \hline
  $\Delta z_{min}/D$   & $1.10 \times 10^{-2}$ & $1.56 \times 10^{-3}$ & $5.37 \times 10^{-4}$\\
  \hline
  $\phi_z$   & $0.9964$ & $0.9926$ & $0.9815$\\ 
  \hline
  $s$  & $20$ & $20$ & $16$
\end{tabular}
 \caption{Grid parameters for stratified turbulent wake test cases. Number of grid points in the $x$ direction is $N_x$, number of grid points in the $z$ direction is $N_z$, and number of grid points in $y$ is $N_y$. Minimum grid spacing in each direction is denoted by $\Delta x/D$, $\Delta y/D$ and $\Delta z_{min}/D$, in nondimensional units based on the sphere diameter $D$. The size of the computational domain is $L_x \times L_y \times L_z = 8D \times 2D \times 1.8D$. In the $z$ direction, the solution is discretized using $N_q=13$ grid points per element in physical space, with varying vertical grid stretching factor between elements $\phi_z$, where the smallest elements are located at the center of the domain. The polynomial expansion of the solution (in $z$) is constant in every element and is selected to be of order $N_p-1=11$. Order of the spectral filter is listed as $s$. The cutoff frequency of the filter is $N_c=0$ and $\alpha= -\log(\epsilon_M)$ (see Appendix A) for all the cases.}
 \label{grid_parameters_wakes}
\end{center}
\end{table}

\subsubsection{Simulation results}

\begin{figure}
    \centering
    \includegraphics[scale=0.33,trim={0cm 0cm 0.1cm 0cm},clip]{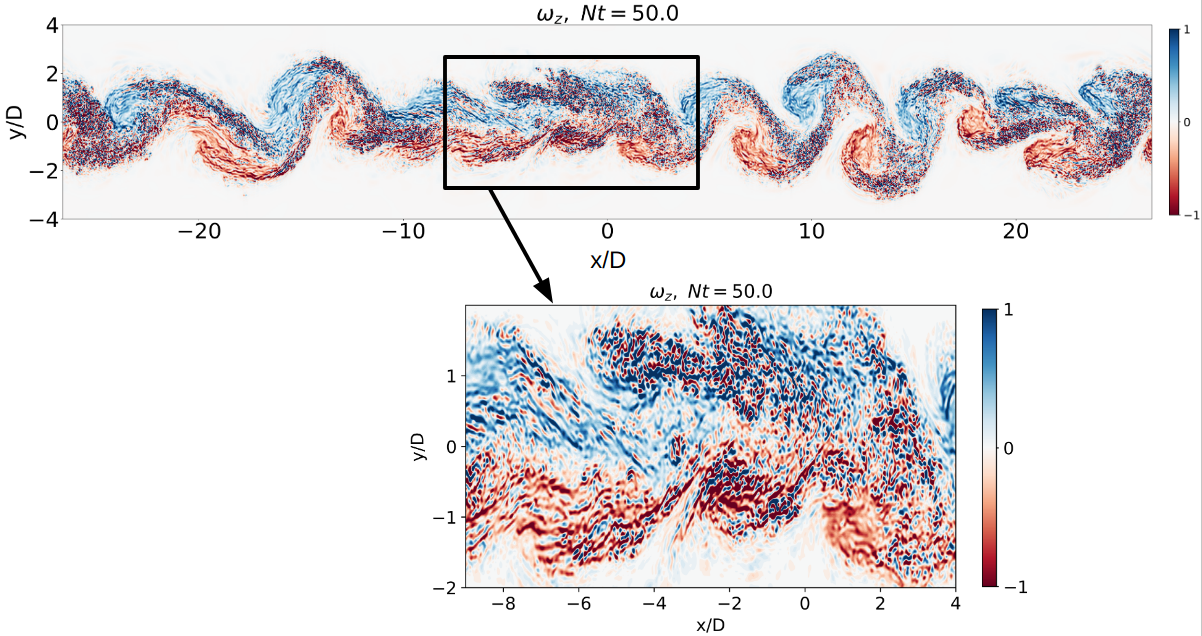}
   \caption{Stratified sphere wake at a body-based Reynolds number $Re = 4 \times 10^5$ and Froude number, $Fr=4$, at $Nt=50$. Contour plots of vertically oriented vorticity, normalized by $N$, at the horizontal center-plane, $z/D=0$. The top plot only shows a fraction of the computational domain in the $y$ direction.}
   \label{figwake400_omegaZ_Nt50}
\end{figure}

\begin{figure}
    \centering
    \includegraphics[scale=0.33,trim={0cm 0cm 0.1cm 0cm},clip]{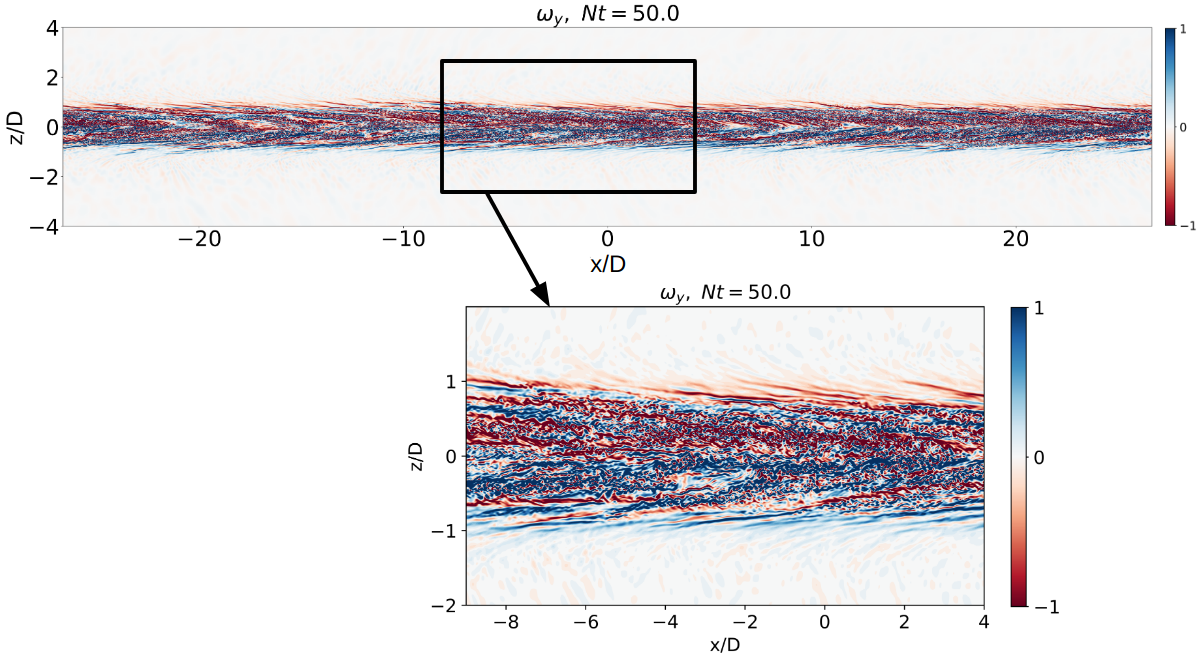}
   \caption{ Stratified sphere wake at a body-based Reynolds number $Re = 4 \times 10^5$ and Froude number, $Fr=4$, at $Nt=50$. Contour plots of span-wise vorticity, normalized by $N$, at the vertical center-plane, $y/D=0$. The body-based Froude number, Fr, is 4. The top plot only shows a fraction of the computational domain in the $z$ direction.}
   \label{figwake400_omegaY_Nt50}
\end{figure}

At a qualitative level, the simulation results exhibit flow dynamics expected for stratified sphere wakes when evaluating quantities such as vorticity and horizontal divergence. 
To give the reader a better sense of the fine spatial resolution we are solving for, Fig.~\ref{figwake400_omegaZ_Nt50} and \ref{figwake400_omegaY_Nt50} offer a view of the total domain length, by showing vorticity contours on the $z/D=0$ and $y/D=0$ planes at $Nt=50$. The zoom-in view in the figures, reveals the presence very fine scale structures in the $Re=4 \times 10^5$ case, compared to the total length of the computational domain. High spatial resolution allows us to represent a very long domain, in the homogeneous direction, while having very fine-scale structures in the flow. 

\begin{figure}
    \centering
    \includegraphics[scale=0.46]{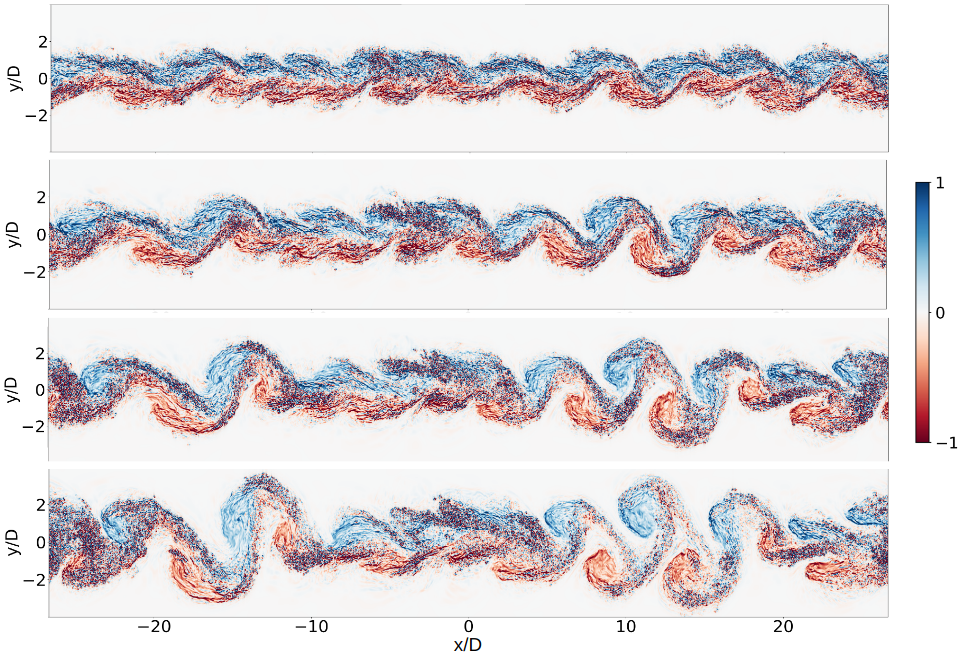}
   \caption{Evolution of stratified sphere wake at a body-based $Re=4\times10^5$. Contour plots of $z$-oriented vorticity, , normalized by $N$, at the horizontal centerplane $z/D=0$. From top to bottom $Nt$:~$30,~40,~50,~60$, where N is the Brunt–Väisälä frequency and $t$ is time. The body-based Froude number, $Fr$, is $4$. Wake width increases over time. Only a fraction of the computational domain is presented in the $y$ direction. In each snapshot, from top to bottom, the maximum values of absolute normalized $z$-oriented vorticity are: $5.98,~5.77,~6.80,~7.09$.}
   \label{figwake400_omegaZ_overtime}
\end{figure}

Two layers of opposite-signed vorticity can be distinguished, with flow structures that look fundamentally different on the horizontal plane ($z/D=0$, see Fig.~\ref{figwake400_omegaZ_Nt50}) and the vertical plane ($y/D=0$, see Fig.~\ref{figwake400_omegaY_Nt50}). The stratified wake growth is not symmetric around the wake centerline, as it is the case for its unstratified counterpart. On the vertical plane, the stratified wake is narrower than in the horizontal, and inclined shear layers start to form due to stratification. As expected, the wake height is constrained by the density differences in the vertical and internal waves begin to emerge. On the contrary, in the horizontal direction the wake spreads at a faster pace and larger-scale vorticity structures, or quasi-horizontal motions, can be distinguished. The development of these quasi-horizontal motions can be observed in more detail in the evolution of the $z-$oriented vorticity field in Fig.~\ref{figwake400_omegaZ_overtime}. For the $Re=4 \times 10^5$ case, large scale vortices are clearly distinguishable after $Nt=50$, along with fine-scale flow structures that are intermittent through the flow. On the contrary, the evolution of the $y-$oriented vorticity does not show the development of larger scale structures; the characteristic length scale of the flow in this direction seems to lock over time. Nevertheless, highly localized regions of shear are clearly established, as illustrated in the evolution of the $y-$oriented vorticity in Fig.~\ref{figwake400_omegaY_Nt50}. As expected, the wake width growth is accompanied by shear instabilities in the vertical direction. The length scale of these flow features, however, changes with $Re$.

\begin{figure}
    \centering
    \includegraphics[scale=0.33]{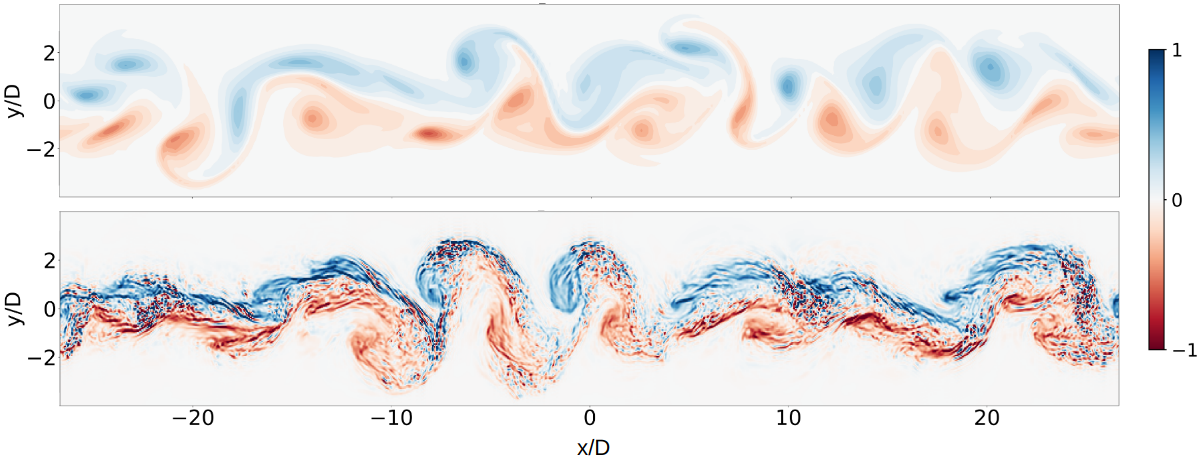}
   \caption{Stratified sphere wake at a body-based $Re=5\times10^3$ (top), $Re=1\times10^5$ (middle) and $Re=4\times10^5$ (bottom) at $Nt=50$. Contour plots of vertically oriented vorticity, normalized by $N$, at the horizontal centerplane. The body-based Froude number, $Fr$, is $4$. Fine scales are clearly intensified at the highest Re. Only a fraction of the computational domain is presented in the $y$ direction.}
   \label{figOmegaZ_Re_Nt50}
\end{figure}

\begin{figure}
    \centering
    \includegraphics[scale=0.33]{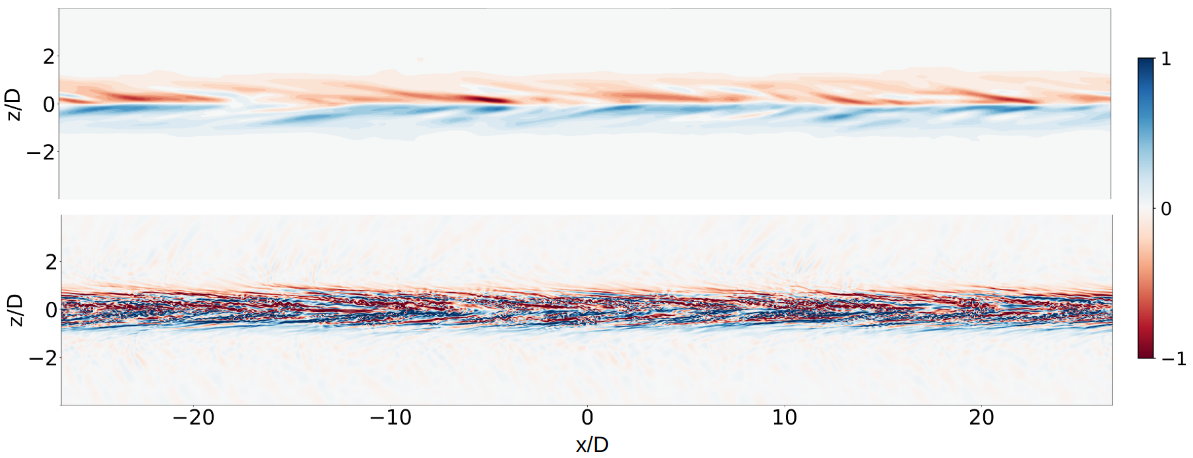}
   \caption{Stratified sphere wake at a body-based $Re=5\times10^3$ (top), $Re=1\times10^5$ (middle) and $Re=4\times10^5$ (bottom) at $Nt=50$. Contour plots of $y-$oriented vorticity, normalized by $N$, at the horizontal centerplane. The body-based Froude number, $Fr$, is $4$. Fine scales and shear are clearly intensified at the highest Re. Only a fraction of the computational domain is presented in the $z$ direction.}
   \label{figOmegaY_Re_Nt50}
\end{figure}

\begin{figure}
    \centering
    \includegraphics[scale=0.33,trim={0cm 0cm 0.1cm 0cm},clip]{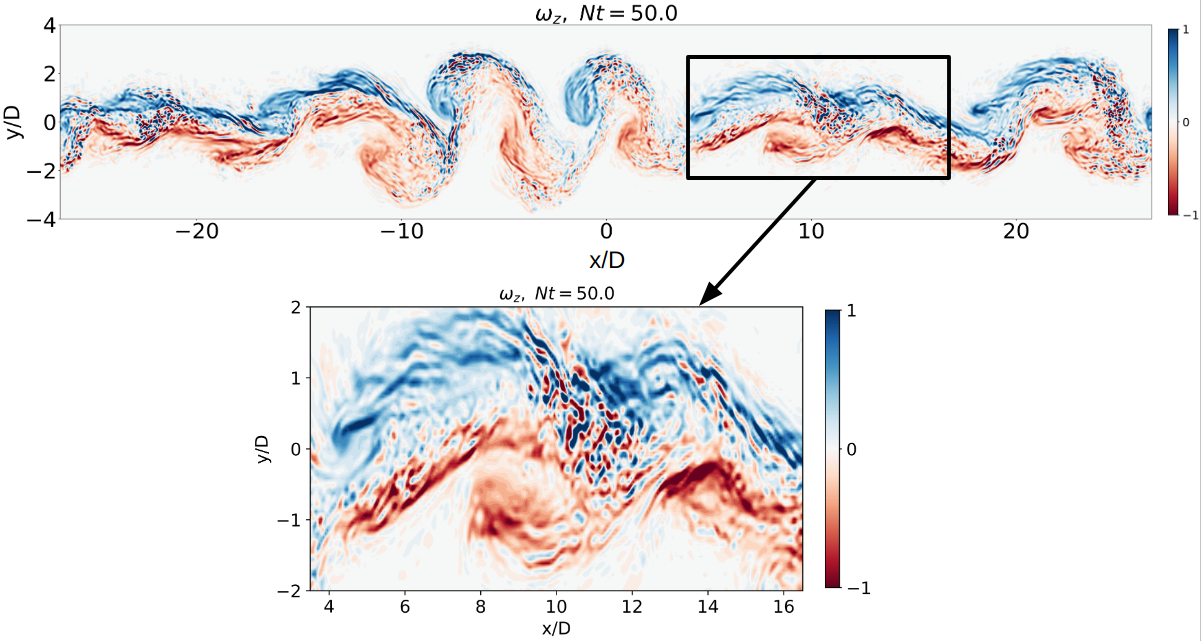}
   \caption{Stratified sphere wake at a body-based Reynolds number $Re = 1 \times 10^5$ and Froude number, $Fr=4$, at $Nt=50$. Contour plots of vertically oriented vorticity, normalized by $N$, at the horizontal center-plane, $z/D=0$. The top plot only shows a fraction of the computational domain in the $y$ direction.}
   \label{figwake100_omegaZ_Nt50}
\end{figure}

\begin{figure}
    \centering
    \includegraphics[scale=0.33,trim={0cm 0cm 0.1cm 0cm},clip]{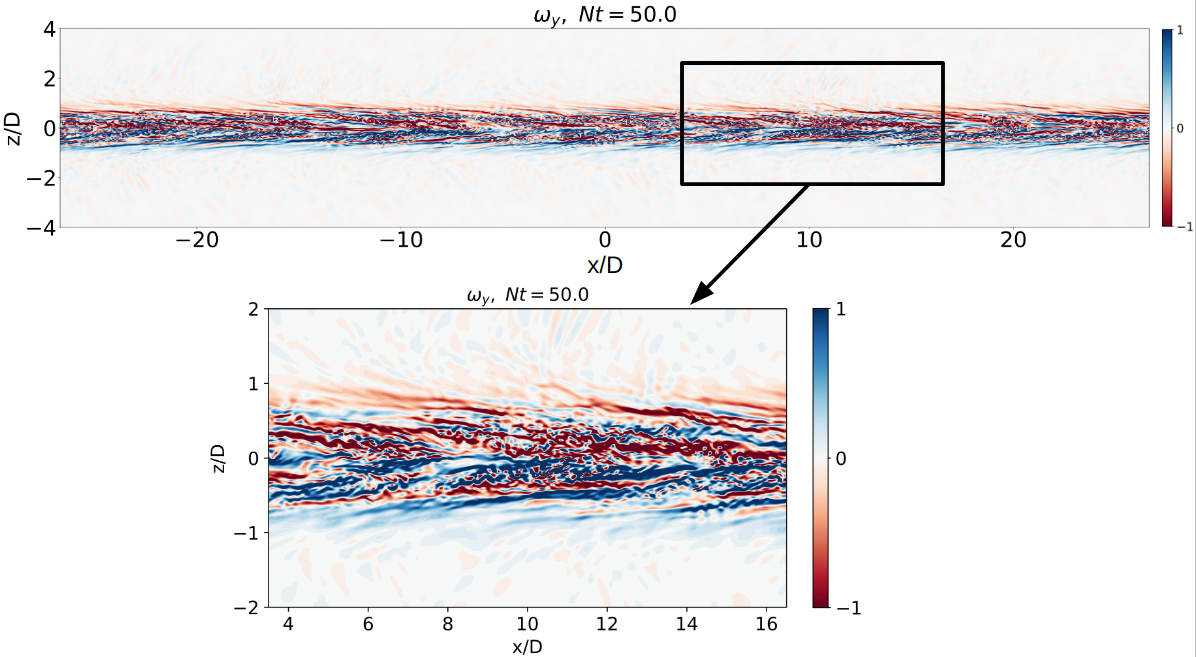}
   \caption{ Stratified sphere wake at a body-based Reynolds number $Re = 1 \times 10^5$ and Froude number, $Fr=4$, at $Nt=50$. Contour plots of span-wise vorticity, normalized by $N$, at the vertical center-plane, $y/D=0$. The body-based Froude number, Fr, is 4. The top plot only shows a fraction of the computational domain in the $z$ direction.}
   \label{figwake100_omegaY_Nt50}
\end{figure}

\begin{figure}
    \centering
    \includegraphics[scale=0.33,trim={0cm 0cm 0.1cm 0cm},clip]{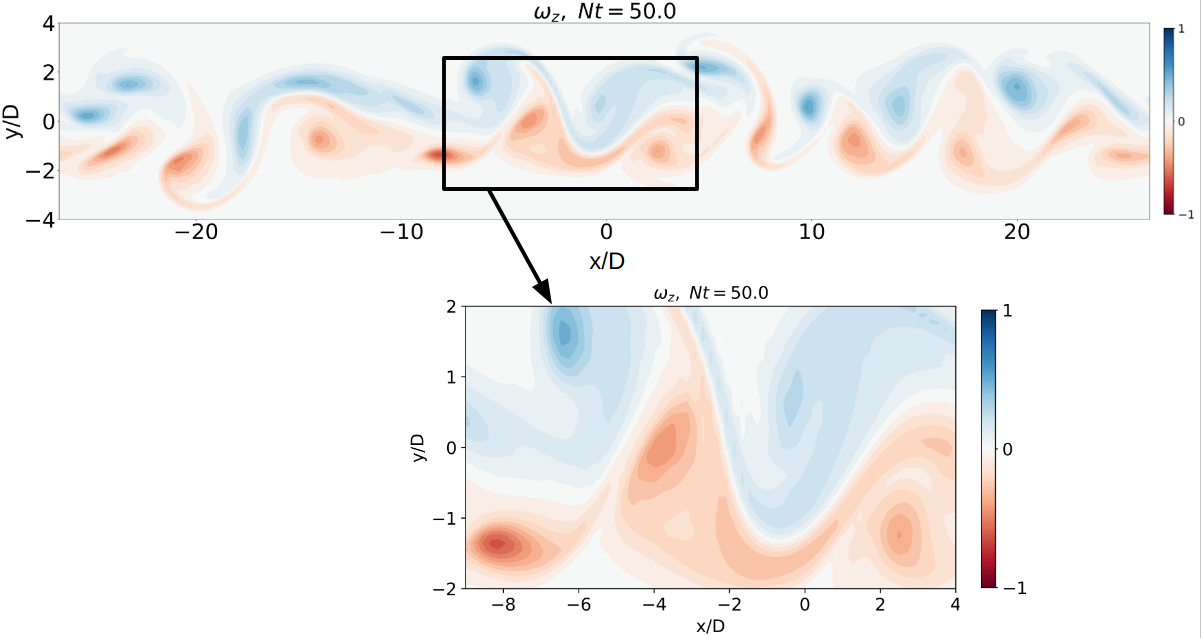}
   \caption{Stratified sphere wake at a body-based Reynolds number $Re = 5 \times 10^3$ and Froude number, $Fr=4$, at $Nt=50$. Contour plots of vertically oriented vorticity, normalized by $N$, at the horizontal center-plane, $z/D=0$. The top plot only shows a fraction of the computational domain in the $y$ direction.}
   \label{figwake5_omegaZ_Nt50}
\end{figure}

\begin{figure}
    \centering
    \includegraphics[scale=0.33,trim={0cm 0cm 0.1cm 0cm},clip]{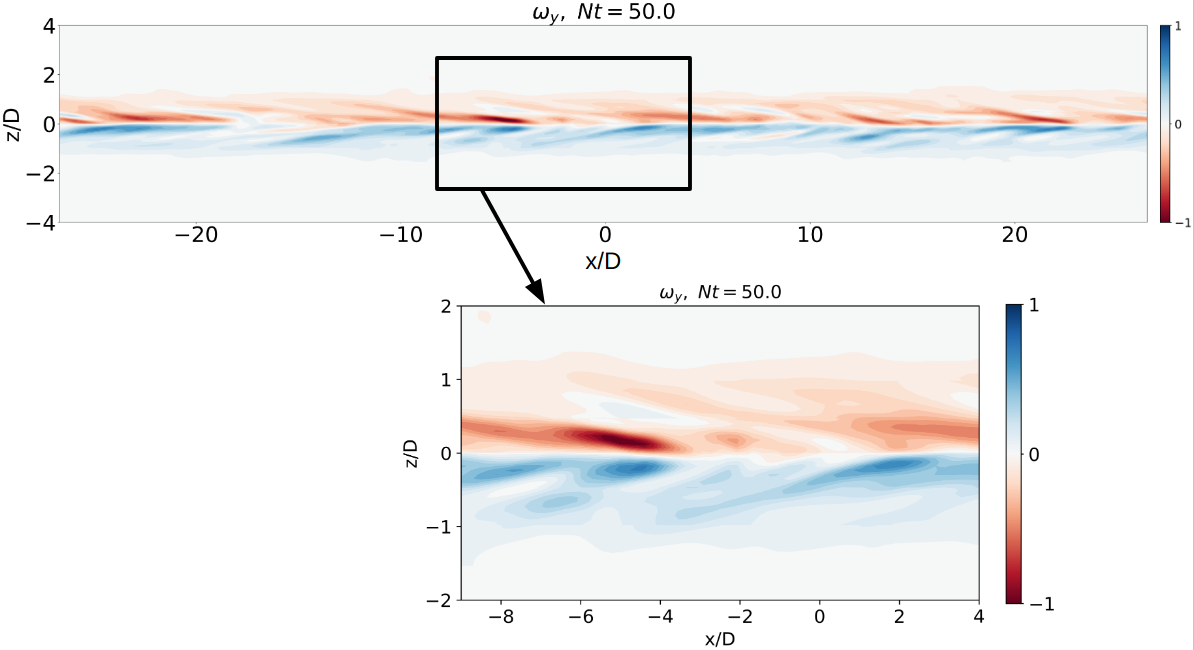}
   \caption{ Stratified sphere wake at a body-based Reynolds number $Re = 5 \times 10^3$ and Froude number, $Fr=4$, at $Nt=50$. Contour plots of span-wise vorticity, normalized by $N$, at the vertical center-plane, $y/D=0$. The body-based Froude number, Fr, is 4. The top plot only shows a fraction of the computational domain in the $z$ direction.}
   \label{figwake5_omegaY_Nt50}
\end{figure}

The difference in the amount of small-scale structures that can be captured, as $Re$ increases, is evident when comparing results across $Re$. Snapshots of vertical and span-wise vorticity are presented in Fig.~\ref{figOmegaZ_Re_Nt50} and Fig.~\ref{figOmegaY_Re_Nt50} for the $Re: 5\times 10^3$ and $1\times 10^5$ cases, at $Nt=50$. Fine scale flow structures and vertical shear are clearly intensified for increasing $Re$. Large scale vortices in the $z-$oriented vorticity contours (see Fig.~\ref{figOmegaZ_Re_Nt50}) are observed for all the $Re$ cases, but small-scale structures and instabilities are present thorough the flow only for the larger $Re$ values. To offer the reader an additional point of comparison, vorticity contours with the same zoom-in view initially presented in Fig.~\ref{figwake400_omegaZ_Nt50} and \ref{figwake400_omegaY_Nt50} for $Re=4\times 10^5$, are now shown for lower $Re$: $Re=5\times 10^3$ in Fig.~\ref{figwake5_omegaZ_Nt50} and \ref{figwake5_omegaY_Nt50}, and $Re=1\times 10^5$ in Fig.~\ref{figwake100_omegaZ_Nt50} and \ref{figwake100_omegaY_Nt50}. Small-scale flow structures and vertical shear can trigger additional instabilities in fully developed turbulence and, practically, more vorticity. As $Re$ increases, the range of possible vertical vorticity values grows, as well as its magnitude. An increase in the amount of vorticity in the flow is expected: the range of possible values of vorticity expands when $Re$ increases, indicating that the flow becomes more vortical, more turbulent. 

\begin{figure}[ht]
    \centering
    \subfloat{\includegraphics[scale=0.32,trim={0cm 0cm 0cm 0.0cm},clip]{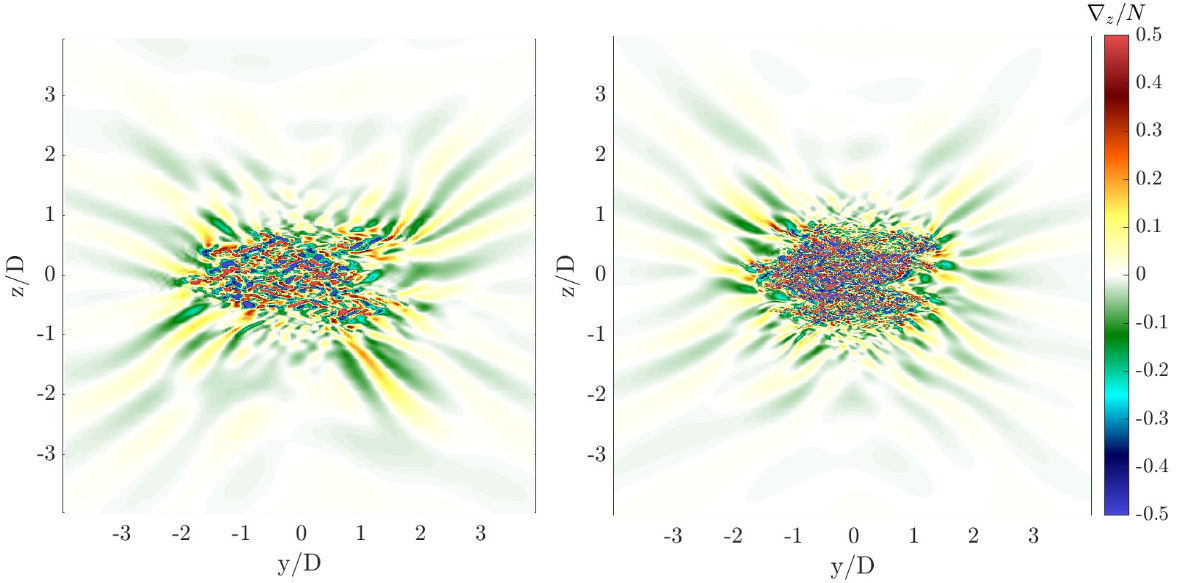}}  
    \caption{Horizontal velocity divergence $\nabla_z$ normalized by buoyancy frequency $N$ at the vertical center plane at $Nt=30$ for a stratified sphere wake. $Re=10^5$ (left column) $Re=4\times 10^5$ (right column). Minimum/maximum contour values correspond only to fraction (about $10\%$) of minimum/maximum values of $\nabla_z/N$ in the flow to facilitate the identification of internal wave radiation. Radiation of IWs is intensified over time. Only a fraction of the computational domain is presented.} 
    \label{fig:IWs_contours}
\end{figure}

For two $Re$ values, Fig.~\ref{fig:IWs_contours} presents a cross-section in the $y$-$z$ plane of the horizontal divergence, $\nabla_z = \partial u / \partial x + \partial v/ \partial y$, as its non-zero values outside the wake core indicate the presence of internal wave motions in the flow. As reported by \cite{DIAMESSIS2005} the latter, the radiation of internal waves at earlier times is evident, and the angle of propagation of internal wave rays is comparable to the approximate value of $45^{\circ}$ from the vertical reported in laboratory data \cite{spedding_2002}.

\begin{figure}
    \centering
    \centering
    \subfloat{\includegraphics[scale=0.26,trim={0.0cm 0cm 0cm 1.35cm},clip] {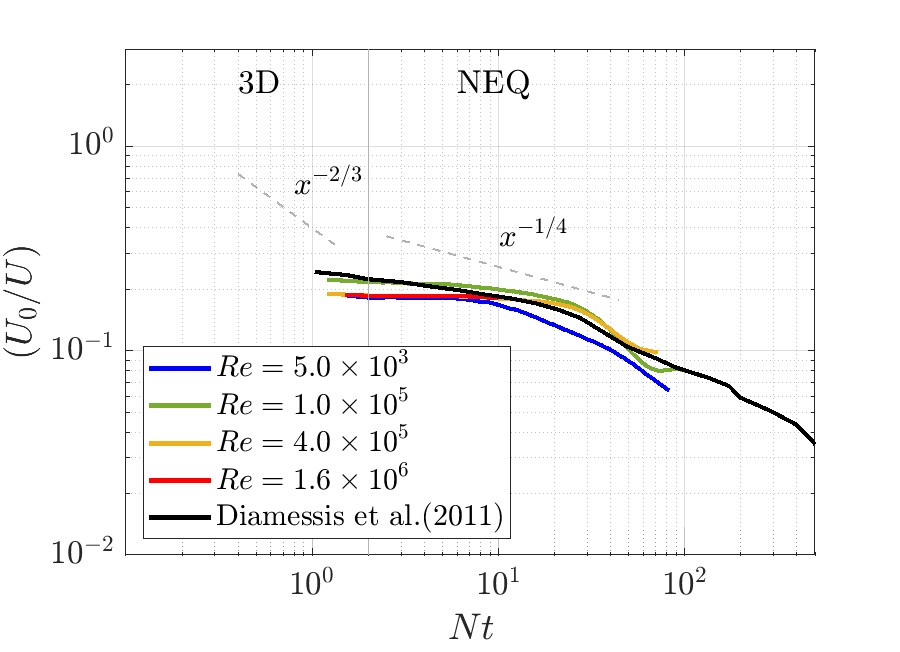}} \\
    \subfloat{\includegraphics[scale=0.26,trim={0.1cm 0cm 1.5cm 0cm},clip]{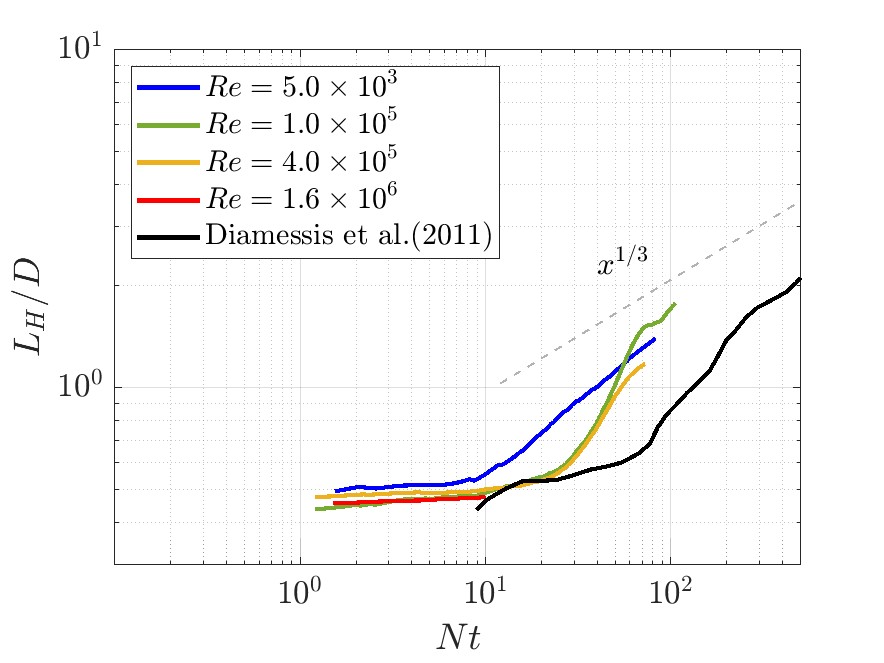}} 
    \subfloat{\includegraphics[scale=0.26,trim={0.1cm 0cm 0cm 0cm},clip]{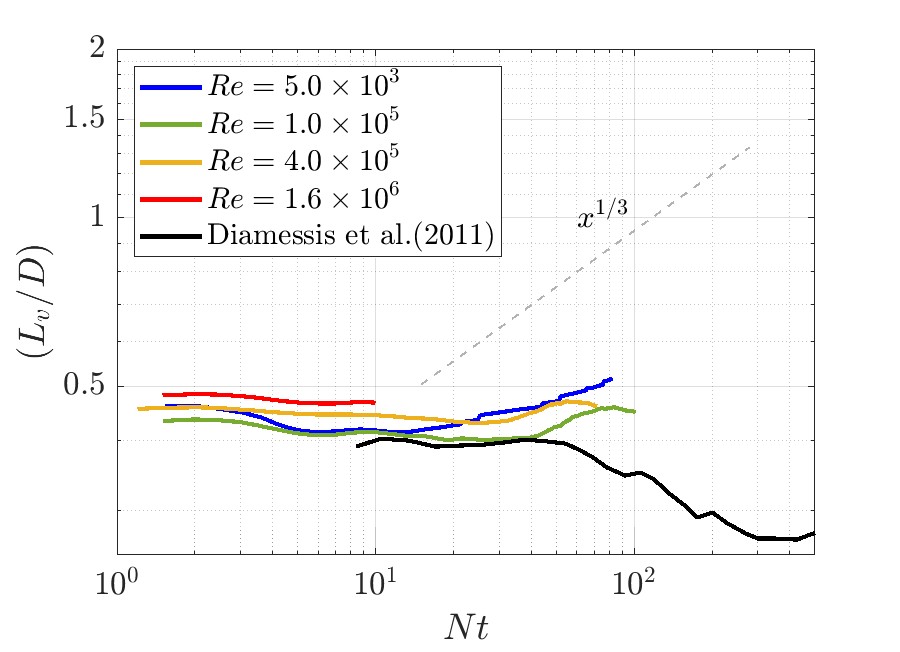}}
\caption{Timeseries of maximum centerline mean velocity $U_0$ (top) and scaled mean velocity proﬁle length scales (bottom) for stratified turbulent wakes with $Fr=4$. Various $Re$ are presented. Classical self-similar scaling laws are indicated in dashed lines. Mean axial velocity $U_0$  and length scales $L_H$ and $L_V$ are obtained using Gaussian fit at the horizontal and vertical center-plane, respectively. Simulated results are compared to previous numerical studies at body-based $Re=5 \times 10^3$ (\cite{DIAMESSIS2005}), and $Re=10^5$ (\cite{diamessis_spedding_domaradzki_2011}).}
\label{stratified_powerLaws}
\end{figure}

As discussed in \cite{spedding_2002, diamessis_spedding_domaradzki_2011}, stratified wakes establish a constant wake height $L_V$, corresponding to the non-equilibrium (NEQ) regime \cite{spedding_1997}, followed by a transition to a $x/D^{3/5}$ growth rate in the quasi-two-dimensional (Q2D) regime. The expected development of $L_V$ during the NEQ regime is reproduced by the flow solver, as shown in bottom-right panel of Fig. \ref{stratified_powerLaws}, where $L_V/D$ is compared against values reported by \cite{diamessis_spedding_domaradzki_2011}. The evolution of wake mean centerline velocity $U_0$ and wake width $L_H$ are also reported and compared in Fig. \ref{stratified_powerLaws}. As in the previous section, the wake velocity and length scales ($U_0$, $L_V$ and $L_H$) were computed by fitting a Gaussian profile to the mean $u(x,r)$ velocity, specifically at the horizontal and vertical center planes ($z/D=0$ and $y/D=0$). For the period of time that was simulated, the results show the expected trend in terms of self-similar decay of characteristic velocity and length scales. 

Performing and analyzing extended numerical simulations, as well as offering a detailed discussion on the wake dynamics for the highest $Re$ cases (including the $Re=1.6 \times 10^6$ simulation) will be part of future research work. The fact that the flow solver is capable of accessing and reproducing properly the mean flow dynamics at the highest body-based $Re$ values that have been previously simulated gives us confidence in the capabilities of the flow solver to simulate high-Reynolds-number stratified wakes.  


\section{Concluding remarks} \label{sec:conclusions}
This study presents a newly developed numerical model for solving the three-dimensional, incompressible Navier-Stokes equations under the Boussinesq approximation. The flow solver utilizes a Fourier pseudo-spectral method in the horizontal direction and a modal spectral element discretization in the vertical direction. It was developed as a continuation of \cite{diamessis_spedding_domaradzki_2011}, and it is specially designed to simulate highly turbulent stratified flow using an implicit large eddy simulation (ILES) strategy. The element discretization gives the user the flexibility of using non-uniform grids in the vertical direction, facilitating the representation of thin horizontal regions of high shear, typical of turbulent stratified flows,  and enabling finer spatial resolution in regions of interest. A boundary-adapted basis of modal type \cite{canuto2007spectral} in combination with static condensation leads to an inexpensive algorithm based on solving small tridiagonal systems. Spurious divergence at element interfaces is strongly contained and we are able to employ lower order polynomials than \cite{diamessis_spedding_domaradzki_2011}. Spectral filtering and over-integration in Lobatto space are applied to address problems of numerical stability caused by under-resolution. This approach enables the examination of high $Re$ dynamics even when full spatial resolution is not achievable due to limitations of computational resources. 

A series of benchmark studies were selected to asses the accuracy of the numerical model. Such test cases include two-dimensional and three-dimensional problems, with and without uniform density profiles. The benchmarks were evaluated in order of complexity, concluding with a turbulent stratiﬁed wake generated by a sphere in a linear stratification profile. The flow solver correctly captures power law exponents and key flow features of the wake in a linear stratification profile, showing good agreement with previous numerical studies (\cite{DIAMESSIS2005, diamessis_spedding_domaradzki_2011}). The implementation of the numerical model meets the design goals and enables the simulation of sphere wakes at the highest body-based $Re$ values that have been previously simulated in the literature, particularly in the context of LES and DNS studies. We are currently performing the simulation of an even higher body-based $Re$, i.e. $Re=1.6 \times 10^6$, using the flow solver (see \cite{Nidiathesis}). The results of the latter exercise will be reported in a forthcoming publication.

Future numerical simulations of sphere wakes will be performed employing this solver, with the purpose of examining turbulent wake dynamics in a broader range of $Re$ and $Fr$ values. Analysis of such results will enable the characterization of the dynamics of the strongly stratified regime in terms of transitional sub-regimes, flow structures, and evolution of velocity and length scales. Investigating internal wave radiation by the wake into the near and far field will also be possible using tools like those utilized by \cite{KrisPaper}. The proposed numerical model is well-suited for investigating additional applications in environmental flow modeling, given its flexibility in local flow resolution in the vertical direction and its ability to simulate high-Reynolds-number flow dynamics in stratified environments. Such applications may include boundary-layer problems, turbulent jets, and mixing processes triggered by internal wave radiation.  

\section*{Acknowledgments}
Funding for this work was provided through the Office of Naval Research by grants N00014-19-1-2101 and N00014-23-1-2172. High-performance computing resources were awarded through the High Performance Computing Modernization Program (HPCMP).


\section*{Apendix A. Stabilization techniques: exponential filtering}
In environmental flow modeling, relying on under-resolved simulations, aliasing effects are commonly addressed by using an exponential filter, as described by \cite{DIAMESSIS2005}. A surrogate of the hyperviscous operator (\cite{MOURA2016401}), the low pass filter function introduces an scale-selective artificial dissipation required to mitigate aliasing driven spectral instabilities. The range of length scales to be filtered and strength of the dissipation is controlled by adjusting the filter parameters, typically the filter order $s$.

As in \cite{DIAMANTOPOULOS2022102065}, the filtering process takes place once every time-step after the non-linear term computation. It involves multiplying each velocity component and the density field by an exponential filtering diagonal matrix $\Lambda$, in each spatial direction separately,in Fourier-Lobatto space. Following \cite{HesthavenWarbuton2007}, the one-dimensional matrix $\Lambda$ is defined as,
\begin{equation}
\Lambda_{ii}=\sigma\Big ( \frac{i-1}{N}\Big),~i=1,\hdots, N.
\end{equation}
where $N-1$ is the order of the Fourier or polynomial expansion in each direction, and the filter function $\sigma$ follows the expression,
\begin{equation}
\sigma(\eta) = \begin{cases} 
      1 & 0\leq \eta \leq \eta_c \\
      exp{\Big( -\alpha \Big( \frac{\eta - \eta_c}{1-\eta_c}\Big )^s \Big )} & \eta_c \leq \eta \leq 1  
\end{cases}
\end{equation}
As in Eq.(5.16) of \cite{HesthavenWarbuton2007}, $\alpha$ is a function of machine epsilon $\varepsilon_M$, $\alpha=\log(\varepsilon_M)$ and $\eta_c=N_c/N$. In the flow solver, $\Lambda$ is applied in each direction ($x$, $y$ and $z$) in its one-dimensional form.  For all the numerical experiments listed in this work, the cutoff $N_c$ is set to be equal to zero, and the structure of the filter function is fully determined by the order $s$, which can take different values in each direction ($x$, $y$ and $z$). 

\bibliographystyle{elsarticle-num} 
\bibliography{myreferences}

\end{document}